\documentclass[prd,showpacs,twocolumn,superscriptaddress,floatfix,nofootinbib,10pt]{revtex4-2}
\usepackage{setspace}
\usepackage[utf8]{inputenc}
\usepackage{float}
\usepackage{amsmath,amssymb,amsfonts,bm}
\usepackage{graphicx}
\usepackage{multirow}
\usepackage[usenames,dvipsnames]{color}
\allowdisplaybreaks
\usepackage[
colorlinks=true,
linkcolor=blue,
breaklinks=true,
urlcolor=blue,
citecolor=blue]{hyperref}

\usepackage{orcidlink}
\usepackage{subfigure}
\usepackage{soul}
\usepackage{xcolor}
\usepackage{physics}
\usepackage{booktabs}
\usepackage{ulem}
\usepackage{tikz-feynman}

\definecolor{green}{rgb}{0,0.6,0}
\newcommand{\mev}{\textrm{ MeV}}

\newcommand{\GXNU}{\affiliation{Department of Physics, Guangxi Normal University, Guilin 541004, China}}

\newcommand{\GXZD}{\affiliation{Guangxi Key Laboratory of Nuclear Physics and Technology, Guangxi Normal University, Guilin 541004, China}}

\newcommand{\CSU}{\affiliation{School of Physics, Central South University, Changsha 410083, China}}

\newcommand{\IMP}{\affiliation{Southern Center for Nuclear-Science Theory (SCNT), Institute of Modern Physics, Chinese Academy of Sciences, Huizhou 516000, China}}

\newcommand{\HIST}{\affiliation{Heavy Ion Science and Technology Key Laboratory, Institute of Modern Physics, Chinese Academy of Sciences, Lanzhou 730000, China}}

\newcommand{\UCAS}{\affiliation{School of Nuclear Sciences and Technology, University of Chinese Academy of Sciences, Beijing 101408, China}}

\newcommand{\IFIC}{\affiliation{Departamento de F\'{\i}sica Te\'orica and IFIC, Centro Mixto Universidad de
Valencia-CSIC Institutos de Investigaci\'on de Paterna, Apartado 22085,
46071 Valencia, Spain}}

\begin{document}
\title{Determination of the binding and $DK$ probability of the $D^{*}_{s0}(2317)$ from the $(\bar{D}\bar K)^-$ mass distributions in $\Lambda_{b}\to \Lambda_{c} (\bar{D}\bar K)^-$ decays}

\begin{abstract}
We study the $\Lambda_{b}\to\Lambda_{c}\bar{D}^{0}K^{-}$ and $\Lambda_{b}\to \Lambda_{c}D^{-}\bar{K}^{0}$ decays which proceed via a Cabibbo and $N_c$ favored process of external emission, and we determine the $\bar{D}^{0}K^{-}$ and $D^{-}\bar{K}^{0}$ mass distributions close to the $\bar{D} \bar{K}$ threshold. For this, we use the tree level contribution plus the rescattering of the meson-meson components, using the extension of the local hidden gauge approach to the charm sector that produces the $D^*_{s0}(2317)$ resonance. We observe a large enhancement of the mass distributions close to threshold due to the presence of this resonance below threshold. Next we undertake the inverse problem of extracting the maximum information on the interaction of the $\bar{D} \bar{K}$ channels from these distributions, and using the resampling method we find that from these data one can obtain precise values of the scattering lengths and effective ranges, the existence of an $I=0$ bound state with a precision of about $4 \;\rm MeV$ in the mass, plus the $\bar{D} \bar{K}$ molecular probability of this state with reasonable precision. Given the fact that the $\Lambda_{b}\to\Lambda_{c}\bar{D}^{0}K^{-}$ decay is already measured by the LHCb collaboration, it is expected that in the next runs with more statistics of the decay, these mass distributions can be measured with precision and the method proposed here can be used to determine the nature of the $D^*_{s0}(2317)$, which is still an issue of debate.   
\end{abstract}

\author{Hai-Peng Li}%
\GXNU%

\author{Wei-Hong Liang\orcidlink{0000-0001-5847-2498}}%
\email[corresponding author: ]{liangwh@gxnu.edu.cn}
\GXNU%
\GXZD%

\author{Chu-Wen Xiao}
\email{xiaochw@gxnu.edu.cn}
\GXNU%
\GXZD%
\CSU%
\author{Ju-Jun Xie}
\email{xiejujun@impcas.ac.cn}
\IMP
\HIST
\UCAS

\author{Eulogio Oset}%
\email{Oset@ific.uv.es}
\GXNU%
\IFIC%

\maketitle

\section{Introduction}\label{sec:Intr}
The $I(J^P) = 0(0^+)$ $D^*_{s0}(2317)$ resonance is a peculiar state. While the quantum numbers allow it to be a standard $c \bar{s}$ state, standard quark models predict $P$-wave masses much higher than the observed one.
The isospin of the state is $I=0$ according to the PDG~\cite{Navas2024}, but the only observed decay mode is $D^+_s\pi^{0}$, which has $I=1$.
This is the mode in which it was first observed~\cite{Aubert2003} and more recent experiments find that the branching ratio to this channel is practically $100\%$~\cite{Ablikim2018}.
Due to this isospin violating mode, the width of the state is extremely small.
Experimentally it is smaller than $3.8 \mev$ and theoretical predictions range from $10 \,\rm keV$ to $100 \,\rm keV$~\cite{hanhart}.
Due to the problems of the standard quark model predictions, the molecular picture, as formed from the  $D^0K^+$, $D^+K^0$, and $D^+_s \eta$ channels, mostly $DK$, is most favored~\cite{vanBeveren:2003kd,Barnes:2003dj,Chen:2004dy,Kolomeitsev:2003ac,Gamermann:2006nm,Guo:2006rp,Guo:2006fu,Yang:2021tvc,Liu:2022dmm}, although other alternatives are also suggested \cite{Godfrey:2003kg,Colangelo:2003vg,Dmitrasinovic:2005gc,Dmitrasinovic:2012zz,Colangelo:2012xi} (see also review paper on ``Heavy Non-$q \bar q$ Mesons"~\cite{hanhart} in the PDG~\cite{Navas2024} and the review of Ref.~\cite{Guo:2023wkv}).
Lattice QCD simulations also support this picture~\cite{Liu:2012zya,Mohler:2013rwa,Lang:2014yfa,Bali:2017pdv,Cheung:2020mql}, and in Ref.~\cite{MartinezTorres:2014kpc}, a detailed analysis of lattice QCD data could quantify the probability of $DK$ in the $D^*_{s0}(2317)$ wave function around $72\%$.
In the molecular picture the state would be bound by about $42 \mev$ with respect to the $D^0K^+$ threshold, quite a large binding compared to other claimed molecular states, like the $T_{cc}(3875)$, which are close to some threshold.
While the support for this kind of molecular structure is solid from theoretical models, 
it would be good to find experimental data that would support the picture.
This would require to involve $DK$ in some reaction, but we know that the state, being bound in this component, cannot decay into this channel.
The closest thing we can think of is to investigate experiments in which $DK$ is produced close to threshold and see which information we can obtain from the inspection of the $DK$ invariant mass distributions.
Intuitively, we can think that the presence of the $DK$ bound state would produce some enhancement in the $DK$ mass distribution close to threshold.
The question is how much, and if one can obtain some information about the $D^*_{s0}(2317)$ from this mass distribution.
This is the purpose of the present work.
One of such reactions could be $\Lambda_b \to \Lambda_c \bar{D}^{0} K^-$, which has been recently investigated by the LHCb collaboration in Ref.~\cite{LHCb:2023eeb}.
For the moment, only the ratio of the decay rate to that of the $\Lambda^{0}_{b}\to\Lambda^{+}_{c}D_s^{-}$ decay, of the order of $20\%$, has been reported, but future upgrades of LHCb should produce enough statistics to access the mass distribution of $\bar{D}^{0}K^-$.
With this perspective we prepare the ground to see the potential of this magnitude to learn about the relationship of the $D^*_{s0}(2317)$ state to the $DK$ component.

To accomplish this task we do the following work: we study the decays $\Lambda_b \to \Lambda_c \bar{D}^0 K^-$ and $\Lambda_b \to \Lambda_c D^- \bar{K}^0$, and evaluate the $\bar{D}^0 K^-$ and $D^- \bar{K}^0$ mass distributions close to their thresholds using the $\bar{D} \bar{K}$ interaction provided by an extension of the local hidden gauge theory \cite{Bando:1987br,Harada:2003jx,Meissner:1987ge,Nagahiro:2008cv}, in which this interaction is driven by the exchange of vector mesons. 
The only unknown of the theory is a regulator of the loops, which is fitted to get the mass of the $D^*_{s0}(2317)$ state at the experimental mass.
Once the theoretical framework is established, we evaluate the mass distributions of $\bar{D}^0 K^-$ and $D^- \bar{K}^0$ close to the threshold, where we observe a large discrepancy with respect to a phase space distribution.

Then comes the second part.
We assume that the experiment has been done and mass distributions like those obtained previously are observed.
Then we tackle the task of extracting the maximum information about the $DK$ interaction.
It is clear that the mass distributions close to the threshold, and far away from the mass of the $D^*_{s0}(2317)$ carry only partial information on the dynamics of the system, but the challenge is to see how much one can learn from it.
For this purpose, we assume certain errors, $5\%$ of the value of the differential width, and tackle the inverse problem of determining the interaction from these pseudodata.
We make fits to the pseudodata and determine the parameters of the theory.
With them we determine the observables, like scattering parameters and the energy of the bound state if it appears.
The important thing, however, is the uncertainties in these observables.
For this, we use the resampling method \cite{Press1992,Efron:1986hys,Albaladejo:2016hae}. 
We anticipate that we can obtain the scattering lengths and effective ranges with a fair precision, and very important, we also obtain a bound state at the right energy, with an uncertainty in the binding energy of about $4 \mev$, much better than the $20 \mev$ uncertainty that one had in this magnitude for the use of the correlation functions \cite{Ikeno:2023ojl}.
We can also determine the molecular probability of the $DK$ component of the $D^*_{s0}(2317)$ with an uncertainty of about $11\%$, also improving the $60\%$ uncertainty obtained for this magnitude from the correlation functions in Ref.~\cite{Ikeno:2023ojl}. 

\section{formalism}\label{sec:form}
\subsection{The $D^*_{s0}(2317)$ state in the chiral unitary approach}
\label{sub:lhg}
We have three coupled channels: $\bar{D}^0K^-$\;(1), $D^-\bar{K}^0$\;(2), $D_s^-\eta$\;(3).
The potential for $i\to j$ transition, using an extension of the local hidden gauge approach \cite{Bando:1987br,Nagahiro:2008cv}, is given by
\begin{equation}
	V_{ij}=C_{ij}\, g^2 \, (p_{1}+p_{3})\cdot(p_{2}+p_{4}),
\end{equation}
which projected in $S$-wave gives
\begin{align}
	\nonumber
	V_{ij}=C_{ij}\, g^2 \,\frac{1}{2}\,\big[3s&-(M^2+m^2+M'^2+m'^2)\\
					   &-\frac{1}{s}(M^2-m^2)(M'^2-m'^2)\big],
\end{align}
with $s$ the c.m. energy squared, and $m,M \; (m', M')$ the masses of the light meson and the charmed meson in channel $i(j)$, respectively.
The $C_{ij}$ coefficients are given by \cite{Ikeno:2023ojl}
\begin{equation}
	C_{ij}=
\begin{pmatrix}
-\dfrac{1}{2}\left(\dfrac{1}{M_{\rho}^2}+\dfrac{1}{M_{\omega}^2}\right)&-\dfrac{1}{M_{\rho}^2}&\dfrac{2}{\sqrt{3}}\dfrac{1}{M_{K^*}^2}\\[4mm]
&-\dfrac{1}{2}\left(\dfrac{1}{M_{\rho}^2}+\dfrac{1}{M_{\omega}^2}\right)&\dfrac{2}{\sqrt{3}}\dfrac{1}{M_{K^*}^2}\\[4mm]
&&0
\end{pmatrix},
\end{equation}
with $M_\rho=775.2 \mev, M_\omega=775.3 \mev$ and $M_{K^*}=893.6 \mev$.
The coupling $g$ of the local hidden gauge approach is given by $g=M_V/2f$, with $M_V=800\; \mev$ and $f=93\; \mev $.

The $D^*_{s0}(2317)$ bound state is found by searching the pole of the scattering matrix
\begin{equation}
	T=[1-VG]^{-1}\, V,
\end{equation}
where $G={\rm diag} (G_i)$, with $G_i$ the loop function of the $\bar D \bar K$ or $D^-_s \eta$ channel.
Following Ref.~\cite{Ikeno:2023ojl}, we have
\begin{eqnarray}\label{eq:Gcut}
G_i (s) = \int_{|{\vec q\,}| < q_{\rm max}} \, \dfrac{\dd^3 q}{(2\pi)^3}  \dfrac{\omega_1 + \omega_2}{2 \,\omega_1 \, \omega_2} \, \dfrac{1}{s-(\omega_1 + \omega_2)^2+i\epsilon},
\end{eqnarray}
with $\omega_1 = \sqrt{M^2 + {\vec{q}}^{\;2}}$, $\omega_2 = \sqrt{m^2 + {\vec{q}}^{\;2}}$ for each of the $i$ channel of the $\bar{D}^0K^-$, $D^-\bar{K}^0$ and $D_s^-\eta$, and $q_{\text{max}}$ is the regulator of the $G$ function that is a measure of the range of the interaction in momentum space~\cite{Gamermann:2009uq,Song:2022yvz}.
The value of $q_{\rm max}$ is chosen here to fix the binding of the $D^*_{s0}(2317)$ state.

The $T$ matrix that we obtain and the usual one in Quantum Mechanics have different normalization.
Following Ref.~\cite{Ikeno:2023ojl}, we find
\begin{equation}
  T = -8 \pi \, \sqrt{s} \, f^{Q M} \approx -8 \pi \,\sqrt{s} \,\dfrac{1}{-\frac{1}{a}+\frac{1}{2}\, r_0 \,k^2-i k},
\label{eq:T8pi}
\end{equation}
with
\begin{equation}
 k=\dfrac{\lambda^{1/2}\left(s, m^2, M^2\right)}{2 \sqrt{s}}.
\end{equation}
Taking into account that
\begin{equation}
 \Im G = -\dfrac{1}{8 \pi \sqrt{s}} \;k,
\end{equation}
we find that
\begin{equation}
 -\dfrac{1}{a}+\dfrac{1}{2}\, r_0 \, k^2 \simeq -8 \pi\, \sqrt{s} \, T^{-1} +i k \,.
\label{eq:ar0}
\end{equation}
Then we can determine the scattering length and effective range for each channel as
\begin{align}
 -\dfrac{1}{a} &=-8 \pi \, \sqrt{s} \,T^{-1} \Big|_{s = s_\text{th}}, \\
 r_0 &=\dfrac{\partial}{\partial k^2}\; 2\, (-8 \pi \, \sqrt{s} \; T^{-1}+ik) \nonumber \\
& =\dfrac{\sqrt{s}}{\mu} \; \dfrac{\partial}{\partial s} \; 2\,(-8 \pi\,  \sqrt{s}\, T^{-1} +i k) \Big|_{s = s_\text{th}} \, ,
\end{align}
where $s_\text{th}$ is the squared of the energy of the system at threshold $(m+M)$ and $\mu$ is the reduced mass of $m$ and $M$.

The couplings are obtained at the pole energy, $\sqrt{s_{0}}$, considering that
\begin{equation}
 T_{ij}  \sim \dfrac{g_i \;g_j}{s-s_0} \,,
\end{equation}
from where
\begin{equation}
 g_1^2=\lim _{s \rightarrow s_0}\left(s-s_0\right)\, T_{11} ; \quad g_j = g_1 \lim _{s\rightarrow s_0} \dfrac{T_{1 j}}{T_{11}},
\end{equation}
and the probabilities of each channel are obtained as \cite{Gamermann:2009uq,Hyodo:2013nka,Hyodo:2011qc}
\begin{equation}
P_i=- g_i^{2} \;\dfrac{\partial G_i}{\partial s} \Big|_{s=s_0} .
\end{equation}

\subsection{The $\bar D^0 K^-, \; D^- \bar K^0$ invariant mass distributions for $\Lambda_b \to \Lambda_c \bar D^0 K^-, \;\Lambda_c D^- \bar K^0$ decays}
\begin{figure}[b] 
\centering
\includegraphics[width=0.35\textwidth]{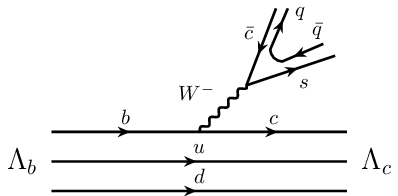}
\caption{Decay mechanism for $\Lambda_{b}\to \bar{c}s\Lambda_{c}$ at the quark level. The $\bar{c}s$ component is hadronized inserting $q\bar{q}$ to obtain two mesons.}
\label{fig:quark}
\end{figure}

In order to evaluate the $\bar{D} \bar{K}$ mass distribution of the $\Lambda_{b}\to\Lambda_{c}^{+}\bar{D}^{0}K^{-}$ and $\Lambda_{b}\to \Lambda_{c}^{+}D^{-}\bar{K}^{0}$, we look at the weak decay process of Fig.~\ref{fig:quark} at the quark level, which proceeds via a Cabibbo favored process with external emission.

After the $\bar{c}s\Lambda_{c}$	production, as depicted in Fig.~\ref{fig:quark}, the $\bar{c}s$ pair is hadronized via 
\begin{equation}\label{eq:scbar}
	s\bar{c}\to\sum_{i}s\bar{q}_{i}q_{i}\bar{c},\qquad q_{i}=u,d,s,c.
\end{equation}
This is easily done recalling that $q_{i}\bar{q}_{j}$ can be written in terms of physical mesons using the SU(4) matrix of pseudoscalar mesons,
\begin{equation}\label{eq:SU4}
	\mathcal{P} =
\scalebox{0.8}{
	\(
	\begin{pmatrix}
    \frac{\eta}{\sqrt{3}}+\frac{\eta^{\prime}}{\sqrt{6}}+\frac{\pi^0}{\sqrt{2}} & \pi^{+} & K^{+} & \bar{D}^0 \\
    \pi^{-} & \frac{\eta}{\sqrt{3}}+\frac{\eta^{\prime}}{\sqrt{6}}-\frac{\pi^0}{\sqrt{2}} & K^0 & D^{-} \\
    K^{-} & \bar{K}^0 & -\frac{\eta}{\sqrt{3}}+\sqrt{\frac{2}{3}} \eta^{\prime} & D_s^{-} \\
    D^0 & D^{+} & D_s^{+} & \eta_c
\end{pmatrix}.
\)
}
\end{equation}
Forgetting the $\eta'$ and $\eta_{c}$ which do not play a role here, we obtain
\begin{align}
	\label{eq:sc}
	\nonumber
	s\bar{c}\to &\sum_{i}\mathcal{P}_{3i}\mathcal{P}_{i4}=(\mathcal{P}^2)_{34}\\
		   &=\bar{D}^{0}K^{-}+D^{-}\bar{K}^{0}-D_{s}^{-}\frac{\eta}{\sqrt{3}}.
\end{align}
We can see that we get the $I=0$ combination of $\bar{D}\bar{K}$, with the isospin multiplets $(\bar{K}^{0}, -K^{-})$, $(\bar{D}^{0}, D^{-})$, as it should be starting with the $I=0$ $s\bar{c}$ pair.
Although apparently we are using SU(4) symmetry with the use of the matrix of Eq.~\eqref{eq:SU4}, in practice we only use the $q\bar q$ content of the mesons, for which the use of Eqs.~\eqref{eq:scbar} and \eqref{eq:SU4} is only convenient but unnecessary to reach Eq.~\eqref{eq:sc}.
Yet, in Eq.~\eqref{eq:scbar} we are giving the same weight to the three $\bar q_i q_i$ components with $u,d,s$ ($q_i =c$ is not used).
Hence, an implicit SU(3) symmetry is assumed (see also Refs.~ \cite{Liang:2014eba,Dias:2019klk} in this respect).

\begin{figure}[b]
\centering
\includegraphics[width=0.4\textwidth]{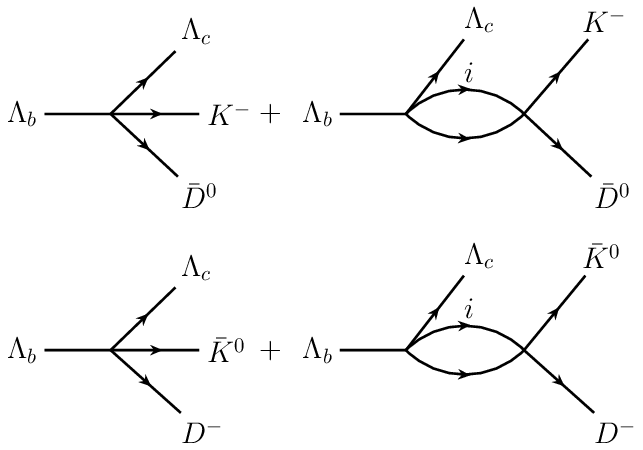}
\caption{Diagrams for $\Lambda_{b}\to\Lambda_{c}\bar{D}^{0}K^{-}\;(D^{-} \bar{K}^{0})$ with tree level and rescattering mechanisms, with $i=\bar{D}^{0}K^{-}, \,D^{-}\bar{K}^{0}, \,D_{s}^{-} \eta$.}
\label{fig:feyn}
\end{figure}

In Eq.~\eqref{eq:sc} we see that both $\bar{D}^{0}K^{-}$ and $D^{-}\bar{K}^{0}$ are produced at tree level with the same weight.
After that we take into account rescattering of these components, and here the $D_{s}^{-}\eta$ component also contributes through the $D_{s}^{-}\eta\to \bar{D}\bar{K}$ transition in coupled channels.
This is depicted in Fig.~\ref{fig:feyn}.

The amplitudes for the two decay processes are then given by
\begin{align}
	&t_{\bar{D}^0 K^-}
	=A\Big\{1+G_{\bar{D}^0 K^-}\; t_{\bar{D}^0K^-,\,\bar{D}^0 K^-}
	\nonumber \\
		 &+G_{D^-\bar{K}^0} \; t_{D^- \bar{K}^0,\bar{D}^0 K^-}
		 -\frac{1}{\sqrt{3}}\, G_{D_s^- \eta} \; t_{D_s^- \eta,\bar{D}^0 K^-}\Big\},
\end{align}
\begin{align}
	&t_{D^- \bar{K}^0}
	=A\Big\{1+G_{\bar{D}^0 K^-}\;t_{\bar{D}^0 K^-,D^- \bar{K}^0} \nonumber\\
   &+G_{D^- \bar{K}^0} \; t_{D^- \bar{K}^0, D^- \bar{K}^0}
   -\frac{1}{\sqrt{3}}\, G_{D_s^- \eta} \; t_{D_s^- \eta,D^- \bar{K}^0}\Big\},
\end{align}
where $A$ is an arbitrary constant.

The partial decay width reads
\begin{equation}\label{eq:Gamma}
	\dfrac{\dd \Gamma}{\dd M_{\text{inv}}}=\dfrac{1}{(2\pi)^3}\;\dfrac{1}{4M_{\Lambda_b}^2}\, p_{\Lambda_c}\, \tilde{p}_{\bar K}\, \overline{\sum}\sum|t|^2,
\end{equation}
with
\begin{equation}
	p_{\Lambda_c}=\dfrac{\lambda^{1/2}\left(M_{\Lambda_b}^2,M_{\Lambda_c}^2,M_{\text{inv}}^2(\bar D \bar K)\right)}{2M_{\Lambda_b}},
\end{equation}
\begin{equation}
	\tilde{p}_{\bar K}=\dfrac{\lambda^{1/2}\left(M_{\text{inv}}^2(\bar D \bar K),M_{\bar K}^2,M_{\bar D}^2\right)}{2M_{\text{inv}}(\bar D \bar K)}.
\end{equation}

From Eq.~\eqref{eq:Gamma}, the invariant mass distributions of $\bar D^0 K^-$ and $D^- \bar K^0$ can be obtained, where there exists a bound state below threshold, as we shall see in Sec.~\ref{sec:res}.
The question is if from these distributions we can determine the binding of this state, and with which precision.
This problem is tackled in the next subsection.

\subsection{Inverse problem}
In order to obtain the properties of the bound state and the scattering magnitudes, we follow a minimum model dependent approach and take the potential matrix, as done in Ref.~\cite{Ikeno:2023ojl},
\begin{equation}\label{eq:V}
	V=\begin{pmatrix}
		V_{11}&V_{12}&V_{13}\\
		&V_{11}&V_{13}\\
		&&0
	\end{pmatrix},
\end{equation}
with
\begin{align}
	\label{eq:v1}
	V_{11}=&V'_{11}+\frac{\alpha}{M_{V}^2}(s-s_0),\\
	V_{12}=&V'_{12}+\frac{\beta}{M_{V}^2}(s-s_0),\\
	\label{eq:v3}
	V_{13}=&V'_{13}+\frac{\gamma}{M_{V}^2}(s-s_0),
\end{align}
where $s_0$ is the energy squared of the $\bar D^0 K^-$ threshold, and the factor $M_{V}^2$ is introduced to make the parameters $\alpha$, $\beta$ and $\gamma$ dimensionless.
Eq.~\eqref{eq:V} implicitly assumes that the potential $V$ is isospin symmetric (isospin can be a bit broken with the unitarization, due to different masses in the same isospin multiplets).
This implies that in the general $V_{ij} (i,j=1,2,3)$ matrix, $V_{11}=V_{22}$ and $V_{13}=V_{23}$ \cite{Ikeno:2023ojl}. 
On the other hand, the energy dependence introduced in Eqs.~\eqref{eq:v1}-\eqref{eq:v3} is done to take into account the possible contributions of other channels, as for instance components of a genuine state \cite{Song:2022yvz,Aceti:2014ala}.

We would like to stress some point here.
In Eqs.~\eqref{eq:V}-\eqref{eq:v3} we state the model that we use to analyze the data. 
Given the fact that all the parameters used here are free, we can call that a model independent analysis, up to the $I=0$ assumption, demanded by the experimental quantum numbers of the $D^*_{s0}(2317)$.
The model used is, thus, in principle, quite different from the one used to generate the data.
It also contains the free parameters $\alpha, \beta, \gamma$ that introduce a possible energy dependence of the potential.
It could well be that when a fit to the pseudodata is done with all this freedom, we realize that no valuable information is obtained for the observables, or in other words, that the uncertainties obtained for the observables are so large that render the method useless.
The resampling method used is a practical and efficient method to obtain these uncertainties, and as we shall see, we can obtain significant values with moderate errors for most of the observables, which makes the method proposed here a valuable tool to analyze the data on mass distributions, providing a strong motivation to the experimental teams to pay attention to these data.

We have now $V'_{11}$, $V'_{12}$, $V'_{13}$, $\alpha$, $\beta$, $\gamma$, $q_{\text{max}}$ and $A$ as free parameters of the theory.
The strategy is to start from the two mass distributions of $\bar D^0 K^-$ and $D^- \bar K^0$, carry a fit to the curves, determine the free parameters of the theory and use them to evaluate the observables. 
For this purpose, we take as pseudodata the values obtained with the local hidden gauge approach.
We take $30$ points equally spaced in energy and assume an ``experimental error" of $5\%$ of the value.
Then, we use the resampling method \cite{Press1992,Efron:1986hys,Albaladejo:2016hae}, generating Gaussian weighted distributions of the centroids of the data and carrying a fit for every sample of these pseudodata.
Every time, the parameters of the interaction are obtained and the observables are evaluated, and after a reasonable number of fits, of the order of $50$, we calculate the average of the observables and their dispersion, which provides the uncertainty by which we can determine these observables from these data.
The procedure is specially recommended when there are many parameters and one expects correlations between them.

\section{Results and discussions}
\label{sec:res}
With the ingredients in Sec.~\ref{sec:form},
the pole position, couplings $g_i$ of the $D^*_{s0}(2317)$ state to different channels, and the probabilities $P_i$ are shown in Table~\ref{tab:Tab1}.
The scattering lengths $a_i$ and effective ranges $r_{0,i}$ are shown in Table~\ref{tab:Tab2}.
Note that the obtained width of the $D^*_{s0}(2317)$ state is zero, since we did not include the isospin violating $D_s \pi$ decay channel, which gives rise to a tiny width.
\begin{table}[b]
\centering
 \caption{Pole position, couplings $g_i$ (in units of MeV) and probabilities $P_i$, with the cutoff $q_{\text{max}} =706 \mev$.}
 \label{tab:Tab1}
\setlength{\tabcolsep}{11pt}
\begin{tabular}{cccc}
\hline
\hline
pole [MeV]& $g_{1}$& $g_{2}$ & $g_{3}$\\
\hline
$2317.85$ & $8182.29$ & $8144.59$ & $-5571.38$\\
\hline\hline
&$P_{1}$&$P_{2}$&$P_{3}$\\
\hline
&$0.34$ & $0.29$ & $0.04$\\
\hline\hline
\end{tabular}
\end{table}

\begin{table}[htbp]
	\centering
	\caption{Scattering lengths $a_i$ and effective ranges $r_{0,i}$. [in units of fm]}
	\label{tab:Tab2}
	\setlength{\tabcolsep}{16pt}
	\begin{tabular}{ccc}
		\hline\hline
		$a_1$& $a_2$& $a_3$\\
		\hline
		$0.70$&$0.51-i0.12$&$0.21-i0.06$\\
		\hline\hline
		$r_{0,1}$& $r_{0,2}$& $r_{0,3}$\\
		\hline
		$-2.25$ & $0.14-i2.41$ & $0.12+i0.11$\\
		\hline\hline
	\end{tabular}
\end{table}

\begin{figure}[t]
    \begin{center}
    \subfigure{
    \includegraphics[width=0.4\textwidth]{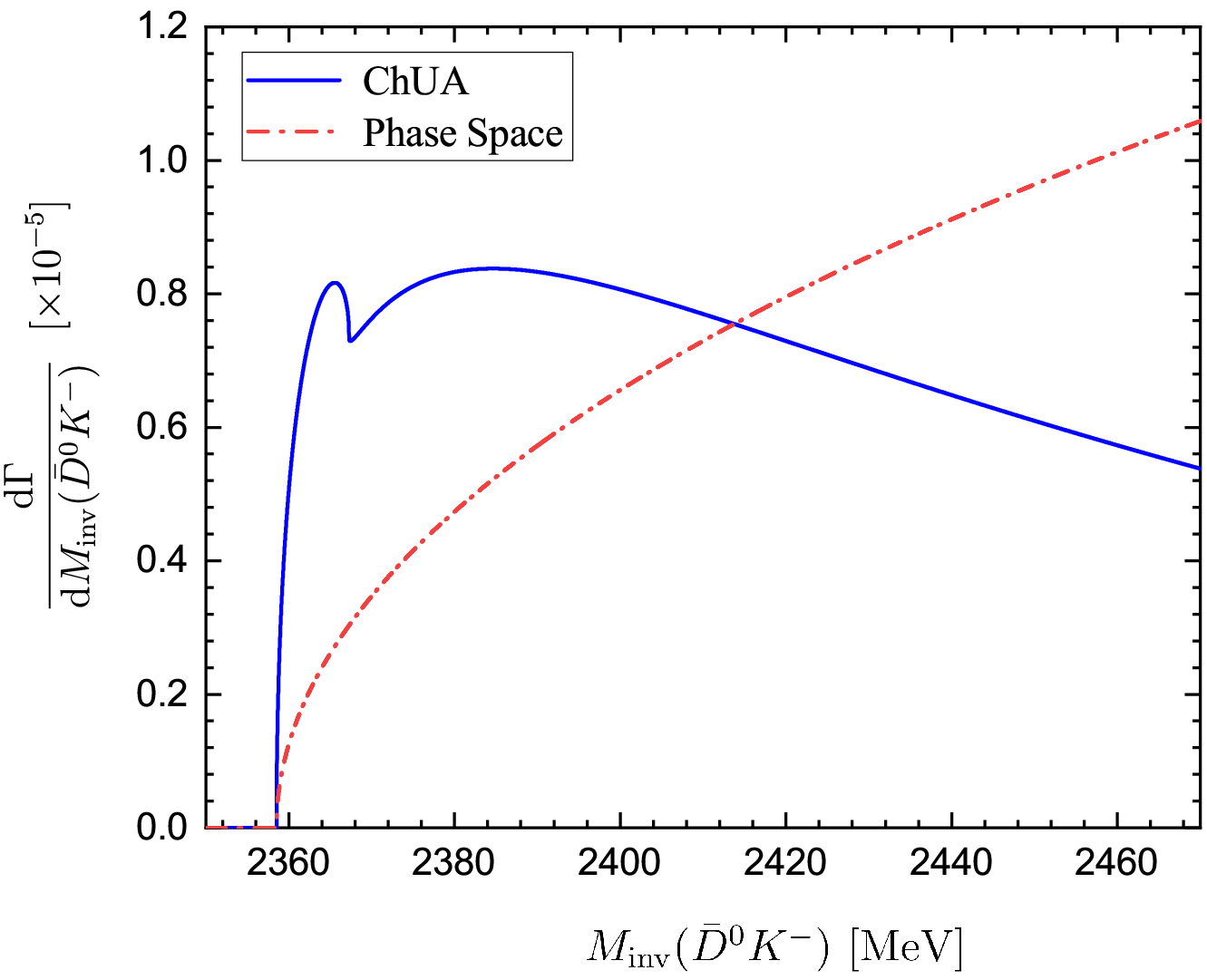}
    }
    \subfigure{
    \includegraphics[width=0.4\textwidth]{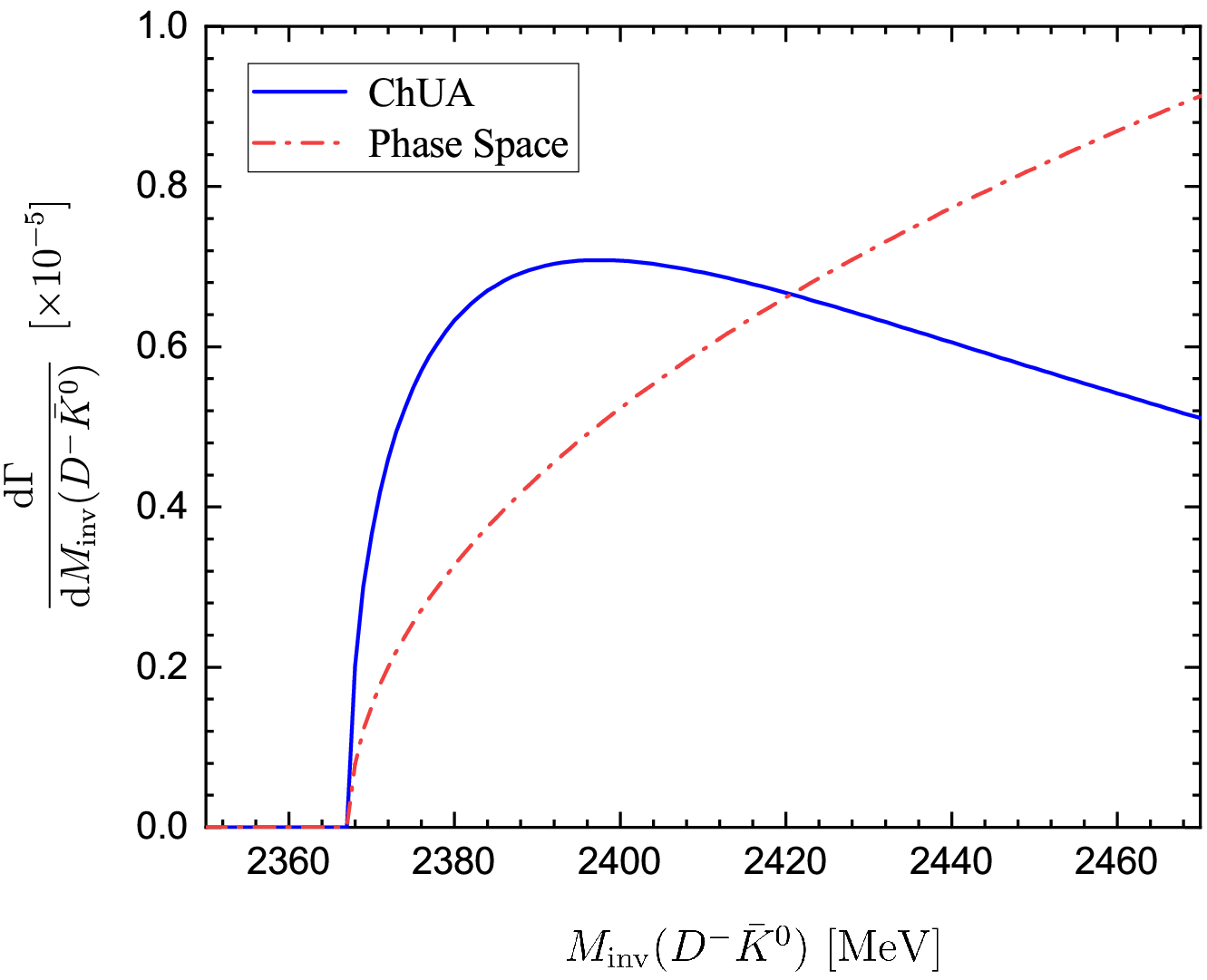}
    }
    \end{center}
    \vspace{-0.7cm}
    \caption{The invariant mass distributions of $\bar{D}^{0}K^{-}$ (up) and $D^{-}\bar{K}^{0}$ (down) using the local hidden gauge approach.}
    \label{Fig:IM}
    \end{figure} 

According to Eq.~\eqref{eq:Gamma}, the obtained $\bar D^0 K^-$ and $D^-\bar K^0$ invariant mass distributions are shown in Fig.~\ref{Fig:IM}, taking $A=1$.
We can see that compared to phase space, with a constant $t$ and normalized to the same area below the curves, the $\bar{D}^{0} K^{-}$ and $D^{-}\bar{K}^{0}$ mass distributions have an important enhancement close to threshold, which is due to the presence of the $D^{*}_{s0}(2317)$ resonance below threshold.
The peak observed in the $\bar{D}^{0}K^{-}$ mass distribution is due to the opening of the $D^{-}\bar{K}^{0}$ channel.

It is thus clear that these invariant mass distributions bear information on the existence of a bound state below threshold.

\begin{figure}[tbph]
\begin{center}
\subfigure{
\includegraphics[width=0.4\textwidth]{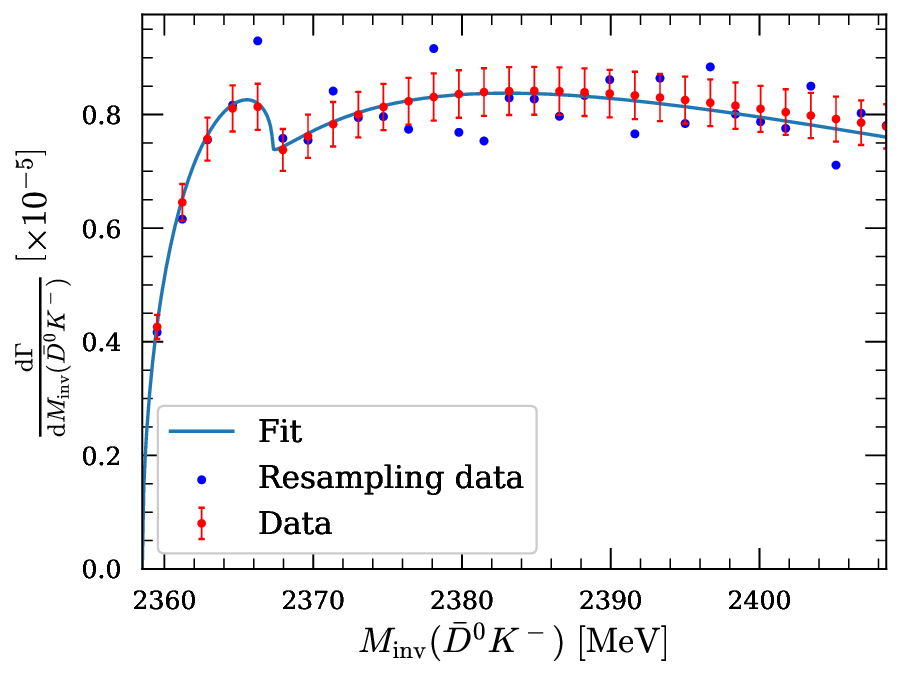}
}
\subfigure{
\includegraphics[width=0.4\textwidth]{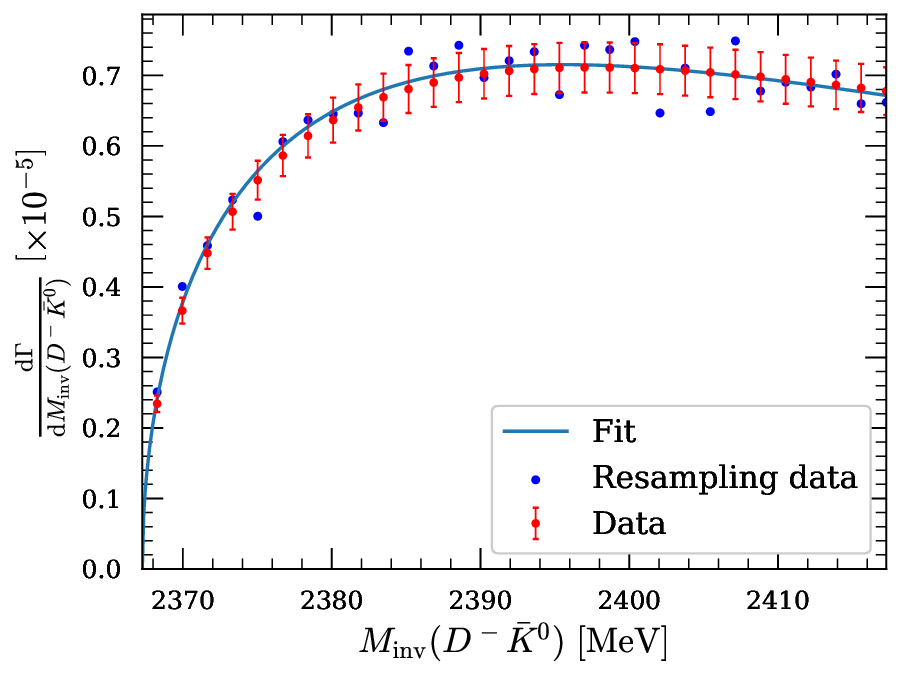}
}
\end{center}
\vspace{-0.7cm}
\caption{Pseudodata taken from the local hidden gauge approach of Subsection~\ref{sub:lhg}. The scattered points are one example of a Gaussian randomly generated centroids of the pseudodata that simulates an actual experimental measurement.}
\label{Fig:fig2}
\end{figure}

Next we discuss the results obtained from the resampling method fits to the data. 
The error of the values and one example of the Gaussian randomly generated values from the resampling are shown in Fig.~\ref{Fig:fig2}.

\begin{table*}[t]
    \centering
    \caption{Values of parameters by fitting the resampled data from thresholds up to $50 \mev$ above thresholds.}
    \label{Tab:tab3}
    \setlength{\tabcolsep}{10pt}
    \begin{tabular}{cccc}
    \hline\hline
    $V_{11}'$& $V'_{12}$ & $V'_{13}$ & $\alpha$\\
    \hline
    $-89.71\pm 31.50$ & $-130.48\pm 27.73$ & $98.15\pm 10.03$ & $38.41\pm 128.72$ \\[3mm]
    \hline\hline
    $\beta$&$\gamma$&$q_{\text{max}}$ [MeV] & $A$\\
    \hline
    $-103.88\pm 120.34$ & $7.28\pm33.75$ & $688.38 \pm 32.87$ & $0.95\pm 0.09$\\
    \hline\hline
    \end{tabular}
\end{table*}
\begin{table*}[t]
    \setlength{\tabcolsep}{11pt}
    \caption{The average and dispersion for pole position and couplings. [in units of MeV]}
    \label{Tab:tab4}
    \begin{tabular}{cccc}
    \hline\hline
    pole& $g_{1}$ & $g_{2}$ & $g_{3}$\\
    \hline
    $2319.32\pm 3.92$ & $8480.92 \pm 612.18$ & $8436.40\pm 578.85$ & $-6594.20\pm 927.89$\\
    \hline\hline
    \end{tabular}
\end{table*}
\begin{table*}[!t]
        \caption{Average value and dispersion of scattering length $a_{i}$ and effective range $r_{0,i}$ for channel $i$. [in units of fm]}
        \label{Tab:far}
    \setlength{\tabcolsep}{14pt}
        \begin{tabular}{ccc}
            \hline\hline
            $a_1$& $a_2$& $a_3$\\
            \hline
            $0.77\pm0.11$&$(0.59\pm0.10)-i(0.11\pm0.03)$&$(0.18\pm0.09)-i(0.05\pm0.02)$\\[3mm]
            \hline\hline
            $r_{0,1}$& $r_{0,2}$& $r_{0,3}$\\
            \hline
            $-1.80\pm0.64$&$(0.42\pm1.09)-i(1.49\pm0.77)$&$-(2.75\pm4.33)-i(1.00\pm1.72)$\\
            \hline\hline
        \end{tabular}
\end{table*}
    
\begin{table}[t]
    \caption{The average and dispersion for the probability.}
    \label{Tab:fp}
    \setlength{\tabcolsep}{13pt}
    \begin{tabular}{ccc}
    \hline
    \hline
    $P_{1}$&$P_{2}$&$P_{3}$\\
    \hline
    $0.38\pm0.06$ & $0.32\pm0.05$ & $0.05\pm0.01$  \\
    \hline
    \end{tabular}
\end{table}

With $V'_{11}$, $V'_{12}$, $V'_{13}$, $\alpha$, $\beta$, $\gamma$, $q_{\text{max}}$ and $A$ as the parameters of the theory, it looks like there are quite many parameters. 
Actually, they are not completely independent, meaning that there could be correlations between them and different sets of parameters could lead to the same solution.
This is actually the case and this is where the resampling method gets its strength.
In each of the fits of the resampling, the fit parameters could be different, but what matters is the values obtained for the observables.   
We take the average of each of them and their dispersion, and the results are shown in Table~\ref{Tab:tab3} for the parameters, in Table~\ref{Tab:tab4} for the pole position and the couplings $g_i$, in Table~\ref{Tab:far} for scattering lengths $a_i$ and effective ranges $r_{0,i}$, and in Table~\ref{Tab:fp} for the probabilities $P_i$.

Since the isospin states of $\bar D \bar K$ are given by
\begin{align}
	\label{eq:iso}
	\ket{\bar{D} \bar{K},I=0}=&\frac{1}{\sqrt{2}}\ket{D^{-}\bar{K}^{0}+\bar{D}^{0}K^{-}},\\
	\ket{\bar{D} \bar{K},I=1}=&\frac{1}{\sqrt{2}}\ket{D^{-}\bar{K}^{0}-\bar{D}^{0}K^{-}},
\end{align}
the potentials for isospin $I=0$, $I=1$ are given by
\begin{align}
	V^{I=0}=&\frac{1}{2}(V_{11}+V_{22}+2V_{12})=V_{11}+V_{12},\\
	V^{I=1}=&\frac{1}{2}(V_{11}+V_{22}-2V_{12})=V_{11}-V_{12}.
\end{align}
We can see that with the results of Table \ref{Tab:tab3}, the $I=0$ interaction is attractive, leading to the $D^{*}_{s0}(2317)$ state, while the interaction in $I=1$ is repulsive and we get no state there.
The state that we get corresponds to $I=0$.
We can also conclude this from inspection of the couplings, $g_{1}$, $g_{2}$ which are practically identical, indicating according to Eq.~\eqref{eq:iso} that we have an $I=0$ state.

We observe that the uncertainties in the parameters are not small, particularly for $\alpha$, $\beta$, $\gamma$.
This is a reflection of the existence of correlations between the parameters, but, as mentioned, what matters is not the values of the parameters but those of the observables and those are obtained with relatively high precision.
Indeed $a_{1}$ for $\bar{D}^{0}K^{-}$ is real and has an uncertainty of $14\%$, $a_{2}$ for $D^{-}\bar{K}^{0}$ is complex (the $\bar{D}^{0}K^{-}$ channel is open for decay) and has uncertainties of about $17\%$ and $27\%$ in the real and imaginary parts respectively.
On the other hand, the $D_{s}^{-} \eta$ channel, which is far away from the $\bar{D} \bar{K}$ thresholds, is also determined but with uncertainties of about $50\%$ and $40\%$ in the real and imaginary parts, respectively.
The effective ranges $r_{0,1}$, $r_{0,2}$ are determined with much bigger uncertainties and $r_{0,3}$ cannot be determined, since the uncertainties are much bigger than the central values.
Very interesting is the results obtained for the pole position, which appear with the right mass and an uncertainty of about $4 \mev$, quite remarkable knowing that the state is bound by $42 \mev$.
The data around the $\bar{D} \bar{K}$ threshold allowed us to find a pole $42 \mev$ below it.
As shown in Table \ref{Tab:fp},
the probabilities are obtained with about $16\%$ accuracy, with $11\%$ errors in the sum of $P_{1}+P_{2}$, summing relative errors in quadrature.
The individual probabilities are $P_{1}=0.38$ for $\bar{D}^{0}K^{-}$ and $P_{2}=0.32$ for $D^{-}\bar{K}^{0}$. $P_{1}+P_{2}\approx 0.7$ indicating that we have largely a bound state of $\bar{D}\bar{K}$.

\begin{figure}[t]
\begin{center}
\subfigure{
\includegraphics[width=0.4\textwidth]{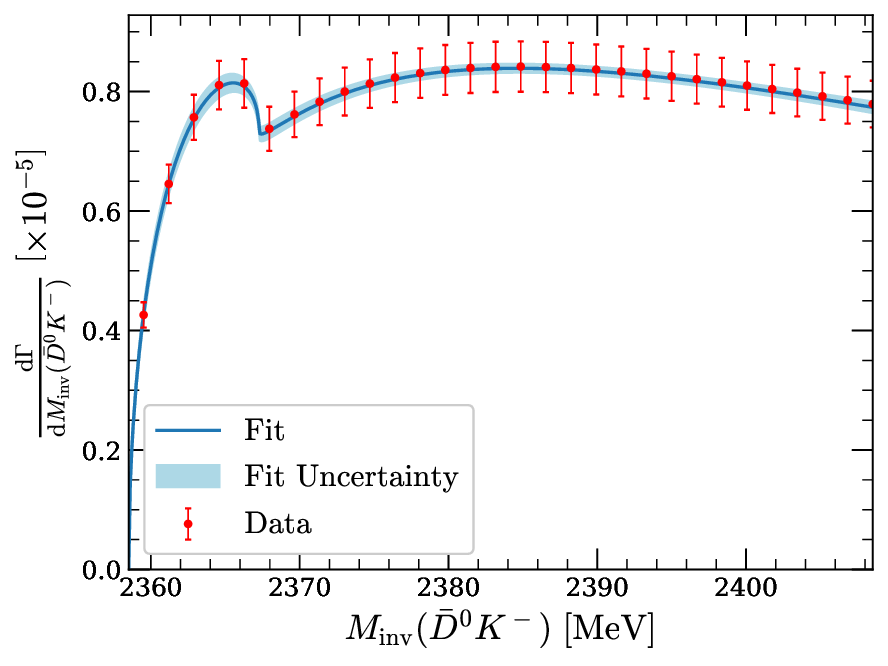}
}
\subfigure{
\includegraphics[width=0.4\textwidth]{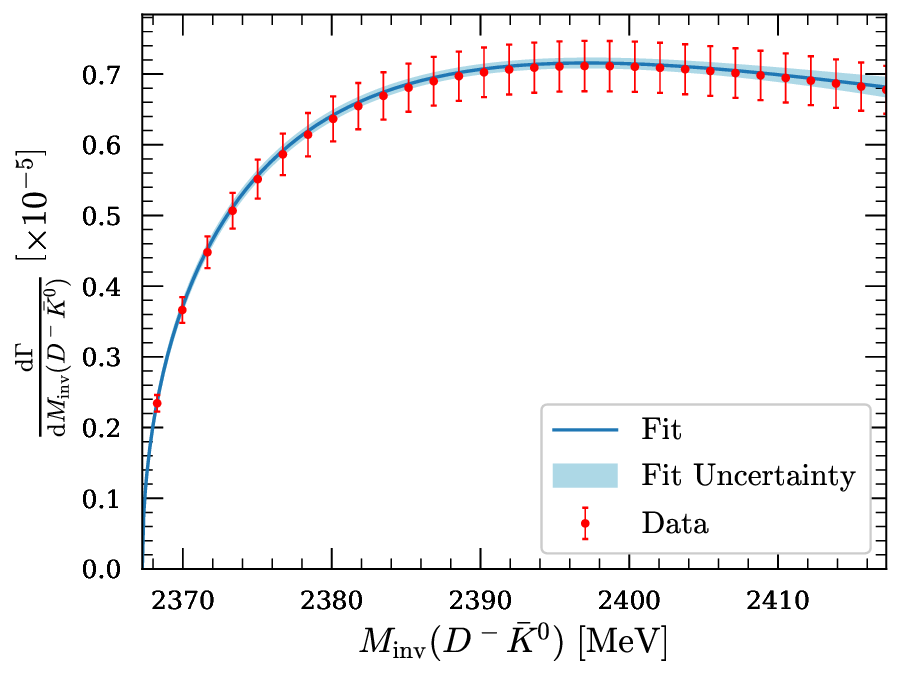}
}
\end{center}
\vspace{-0.7cm}
\caption{Results of the fit from the resampling procedure with the error band (see the text). }
\label{Fig:fig3}
\end{figure}

In Fig.~\ref{Fig:fig3} we also show the results of the resampling with a narrow band for the uncertainty.
Altogether we see that we could extract a large amount of information from the analysis of the $\bar{D}^{0}K^{-}$ and $D^{-}\bar{K}^{0}$ invariant mass distributions close to threshold from the $\Lambda_{b}\to\Lambda_{c}\bar{D}^{0}K^{-}$ and $\Lambda_{b}\to\Lambda_{c}D^{-}\bar{K}^{0}$ decays.

\section{Conclusions}
We have studied the $\Lambda_b \to \Lambda_c \bar{D}^0 K^-$ and $\Lambda_b \to \Lambda_c D^- \bar{K}^0$ decays and shown that they proceed via a weak decay mechanism of external emission which is Cabibbo and $N_c$ favored.   
After the quark level decay is identified, we proceed to hadronize the produced $s\bar c$ pair into two mesons and find that $\bar{D}^0 K^-$ and $D^- \bar{K}^0$ are produced with equal weight, together with $D_s^- \eta$.
The next step consist in allowing these components to undergo final state interaction (rescattering) to have at the end the desired final state. 
The interaction of the $\bar{D}^0 K^-$, $D^- \bar{K}^0$ and $D_s^- \eta$ channels is constructed using an extension of the local hidden gauge approach that gives rise to the $D^*_{s0}(2317)$ state. 
The rescattering procedure leads to $\bar{D}^0 K^-$ and $D^- \bar{K}^0$ mass distributions, peaking sharply at threshold, that differ appreciably from phase space. 
This is due to the presence of the $D^*_{s0}(2317)$ resonance which is bound by $42 \mev$ but makes itself felt at the threshold of the $\bar{D} \bar{K}$.

The next step consist in assuming that the $\bar{D}^0 K^-$ and $D^- \bar{K}^0$ mass distributions are measured in a certain span of invariant masses (so far only branching ratios are measured for the first decay), and we try to see what information can be obtained from them using a minimum model dependent method to analyze the data. 
For this purpose, we assume that the data are those given by the local hidden gauge approach and carry a fit to these data. 
We use then a general transition potential between the channels, with the matrix elements of the potential considered as free parameters. 
We make these matrix elements energy dependent to take into account possible missing channels or the existence of a genuine state of nonmolecular nature, such that, if the $D^*_{s0}(2317)$ state were not a molecular state, the fit could take that into account. 
We should note that in the case that the $D^*_{s0}(2317)$ state were not linked to the $\bar{D} \bar{K}$ components we should not expect any enhancement in the $\bar{D}^0 K^-$ and $D^- \bar{K}^0$ mass distributions at threshold due to the presence of that state. 
Hence, the data will tell us if this is the case or not. 
For the moment we assumed that the data are those of the local hidden gauge approach that generates the $D^*_{s0}(2317)$ state and we carry a fit to the data.

The procedure of the fit is done using the resampling method, specially advised when there are correlations between the parameters of the theory, as we have in the present case.
We generate randomly gaussian points of the mass distributions and carry a fit to these pseudodata many times obtaining the parameters of the theory, 
and then we evaluate the scattering lengths and effective ranges plus the pole position (if a pole appears, which is the case) for each resampled set of pseudodata.
Finally the average of all these observables and their dispersion are evaluated.
We find that these data allow us to obtain good values of the scattering lengths with small uncertainties, and also values for the effective ranges with larger uncertainties.
The most remarkable fact is that we can obtain a pole corresponding to a state of $I=0$ at the right position of the $D^*_{s0}(2317)$ and with uncertainty of only $4 \mev$ in the mass, assuming about $5\%$ error in the strength of the mass distributions.
We are also able to determine the molecular probability of the $\bar{D} \bar{K}$ component with an uncertainty of about $11\%$.
These results are telling us the great potential that these data have in order to determine scattering parameters of the $\bar{D}^0 K^-$ and $D^- \bar{K}^0$ pairs and the molecular nature of the $D^*_{s0}(2317)$. 
In view of these excellent results, we can only encourage the measurement of these mass distributions, only one step forward from present measurements on these decays at LHCb, that could be accomplished in the next planned run in the CERN facility.

\section*{Acknowledgments}
We would like to acknowledge first results from Natsumi Ikeno showing the mass distributions of $D^-\bar K^0$ and $\bar D^0 K^-$ at threshold, as reported in this paper.
This work is partly supported by the National Natural Science Foundation of China (NSFC) under Grants No. 12365019, No. 11975083, No. 12075288, No. 12435007, and No. 12361141819,
and by the Central Government Guidance Funds for Local Scientific and Technological Development, China (No. Guike ZY22096024), 
and by the Natural Science Foundation of Guangxi province under Grant No. 2023JJA110076, and partly by the Natural Science Foundation of Changsha under Grant No. kq2208257 and the Natural Science Foundation of Hunan province under Grant No. 2023JJ30647 (CWX).
This work is supported partly by the Spanish Ministerio de Ciencia e Innovaci\'on (MICINN) under contracts PID2020-112777GB-I00, PID2023-147458NB-C21 and CEX2023-001292-S; by Generalitat Valenciana under contracts PROMETEO/2020/023 and CIPROM/2023/59.

\bibliographystyle{a}
\bibliography{refs}

\begin{thebibliography}{41}%
\makeatletter
\providecommand \@ifxundefined [1]{%
 \@ifx{#1\undefined}
}%
\providecommand \@ifnum [1]{%
 \ifnum #1\expandafter \@firstoftwo
 \else \expandafter \@secondoftwo
 \fi
}%
\providecommand \@ifx [1]{%
 \ifx #1\expandafter \@firstoftwo
 \else \expandafter \@secondoftwo
 \fi
}%
\providecommand \natexlab [1]{#1}%
\providecommand \enquote  [1]{``#1''}%
\providecommand \bibnamefont  [1]{#1}%
\providecommand \bibfnamefont [1]{#1}%
\providecommand \citenamefont [1]{#1}%
\providecommand \href@noop [0]{\@secondoftwo}%
\providecommand \href [0]{\begingroup \@sanitize@url \@href}%
\providecommand \@href[1]{\@@startlink{#1}\@@href}%
\providecommand \@@href[1]{\endgroup#1\@@endlink}%
\providecommand \@sanitize@url [0]{\catcode `\\12\catcode `\$12\catcode `\&12\catcode `\#12\catcode `\^12\catcode `\_12\catcode `\%12\relax}%
\providecommand \@@startlink[1]{}%
\providecommand \@@endlink[0]{}%
\providecommand \url  [0]{\begingroup\@sanitize@url \@url }%
\providecommand \@url [1]{\endgroup\@href {#1}{\urlprefix }}%
\providecommand \urlprefix  [0]{URL }%
\providecommand \Eprint [0]{\href }%
\providecommand \doibase [0]{https://doi.org/}%
\providecommand \selectlanguage [0]{\@gobble}%
\providecommand \bibinfo  [0]{\@secondoftwo}%
\providecommand \bibfield  [0]{\@secondoftwo}%
\providecommand \translation [1]{[#1]}%
\providecommand \BibitemOpen [0]{}%
\providecommand \bibitemStop [0]{}%
\providecommand \bibitemNoStop [0]{.\EOS\space}%
\providecommand \EOS [0]{\spacefactor3000\relax}%
\providecommand \BibitemShut  [1]{\csname bibitem#1\endcsname}%
\let\auto@bib@innerbib\@empty
\bibitem [{\citenamefont {Navas}\ {\it et~al.}(2024)\citenamefont {Navas} {\it et~al.}}]{Navas2024}%
  \BibitemOpen
  \bibfield  {author} {\bibinfo {author} {\bibfnamefont {S.}~\bibnamefont {Navas}} {\it et~al.} (\bibinfo {collaboration} {Particle Data Group}),\ }\bibinfo {title} {{Review of particle physics}},\ \href {https://doi.org/10.1103/PhysRevD.110.030001} {\bibfield  {journal} {\bibinfo  {journal} {Phys. Rev. D}\ }\textbf {\bibinfo {volume} {110}},\ \bibinfo {pages} {030001} (\bibinfo {year} {2024})}\BibitemShut {NoStop}%
\bibitem [{\citenamefont {Aubert}\ {\it et~al.}(2003)\citenamefont {Aubert} {\it et~al.}}]{Aubert2003}%
  \BibitemOpen
  \bibfield  {author} {\bibinfo {author} {\bibfnamefont {B.}~\bibnamefont {Aubert}} {\it et~al.} (\bibinfo {collaboration} {BaBar}),\ }\bibinfo {title} {{Observation of a narrow meson decaying to $D_s^+ \pi^0$ at a mass of 2.32-GeV/c$^2$}},\ \href {https://doi.org/10.1103/PhysRevLett.90.242001} {\bibfield  {journal} {\bibinfo  {journal} {Phys. Rev. Lett.}\ }\textbf {\bibinfo {volume} {90}},\ \bibinfo {pages} {242001} (\bibinfo {year} {2003})},\ \Eprint {https://arxiv.org/abs/hep-ex/0304021} {arXiv:hep-ex/0304021} \BibitemShut {NoStop}%
\bibitem [{\citenamefont {Ablikim}\ {\it et~al.}(2018)\citenamefont {Ablikim} {\it et~al.}}]{Ablikim2018}%
  \BibitemOpen
  \bibfield  {author} {\bibinfo {author} {\bibfnamefont {M.}~\bibnamefont {Ablikim}} {\it et~al.} (\bibinfo {collaboration} {BESIII}),\ }\bibinfo {title} {{Measurement of the absolute branching fraction of $D_{s0}^{*\pm}(2317)\to \pi^0 D_{s}^{\pm}$}},\ \href {https://doi.org/10.1103/PhysRevD.97.051103} {\bibfield  {journal} {\bibinfo  {journal} {Phys. Rev. D}\ }\textbf {\bibinfo {volume} {97}},\ \bibinfo {pages} {051103} (\bibinfo {year} {2018})},\ \Eprint {https://arxiv.org/abs/1711.08293} {arXiv:1711.08293 [hep-ex]} \BibitemShut {NoStop}%
\bibitem [{\citenamefont {Gutsche}\ {\it et~al.}()\citenamefont {Gutsche}, \citenamefont {Hanhart},\ and\ \citenamefont {Mitchell}}]{hanhart}%
  \BibitemOpen
  \bibfield  {author} {\bibinfo {author} {\bibfnamefont {T.}~\bibnamefont {Gutsche}}, \bibinfo {author} {\bibfnamefont {C.}~\bibnamefont {Hanhart}},\ and\ \bibinfo {author} {\bibfnamefont {R.}~\bibnamefont {Mitchell}},\ }\href@noop {} {\bibinfo {title} {{Heavy Non-$q\bar q$ Mesons, review paper in the PDG \cite{Navas2024}}}}\BibitemShut {NoStop}%
\bibitem [{\citenamefont {van Beveren}\ and\ \citenamefont {Rupp}(2003)}]{vanBeveren:2003kd}%
  \BibitemOpen
  \bibfield  {author} {\bibinfo {author} {\bibfnamefont {E.}~\bibnamefont {van Beveren}}\ and\ \bibinfo {author} {\bibfnamefont {G.}~\bibnamefont {Rupp}},\ }\bibinfo {title} {{Observed $D_s(2317)$ and tentative $D(2100\text{--}2300)$ as the charmed cousins of the light scalar nonet}},\ \href {https://doi.org/10.1103/PhysRevLett.91.012003} {\bibfield  {journal} {\bibinfo  {journal} {Phys. Rev. Lett.}\ }\textbf {\bibinfo {volume} {91}},\ \bibinfo {pages} {012003} (\bibinfo {year} {2003})},\ \Eprint {https://arxiv.org/abs/hep-ph/0305035} {arXiv:hep-ph/0305035} \BibitemShut {NoStop}%
\bibitem [{\citenamefont {Barnes}\ {\it et~al.}(2003)\citenamefont {Barnes}, \citenamefont {Close},\ and\ \citenamefont {Lipkin}}]{Barnes:2003dj}%
  \BibitemOpen
  \bibfield  {author} {\bibinfo {author} {\bibfnamefont {T.}~\bibnamefont {Barnes}}, \bibinfo {author} {\bibfnamefont {F.~E.}\ \bibnamefont {Close}},\ and\ \bibinfo {author} {\bibfnamefont {H.~J.}\ \bibnamefont {Lipkin}},\ }\bibinfo {title} {{Implications of a $DK$ molecule at 2.32-GeV}},\ \href {https://doi.org/10.1103/PhysRevD.68.054006} {\bibfield  {journal} {\bibinfo  {journal} {Phys. Rev. D}\ }\textbf {\bibinfo {volume} {68}},\ \bibinfo {pages} {054006} (\bibinfo {year} {2003})},\ \Eprint {https://arxiv.org/abs/hep-ph/0305025} {arXiv:hep-ph/0305025} \BibitemShut {NoStop}%
\bibitem [{\citenamefont {Chen}\ and\ \citenamefont {Li}(2004)}]{Chen:2004dy}%
  \BibitemOpen
  \bibfield  {author} {\bibinfo {author} {\bibfnamefont {Y.-Q.}\ \bibnamefont {Chen}}\ and\ \bibinfo {author} {\bibfnamefont {X.-Q.}\ \bibnamefont {Li}},\ }\bibinfo {title} {{A Comprehensive four-quark interpretation of $D_{(s)}(2317), D_{(s)}(2457)$ and $D_{(s)}(2632)$}},\ \href {https://doi.org/10.1103/PhysRevLett.93.232001} {\bibfield  {journal} {\bibinfo  {journal} {Phys. Rev. Lett.}\ }\textbf {\bibinfo {volume} {93}},\ \bibinfo {pages} {232001} (\bibinfo {year} {2004})},\ \Eprint {https://arxiv.org/abs/hep-ph/0407062} {arXiv:hep-ph/0407062} \BibitemShut {NoStop}%
\bibitem [{\citenamefont {Kolomeitsev}\ and\ \citenamefont {Lutz}(2004)}]{Kolomeitsev:2003ac}%
  \BibitemOpen
  \bibfield  {author} {\bibinfo {author} {\bibfnamefont {E.~E.}\ \bibnamefont {Kolomeitsev}}\ and\ \bibinfo {author} {\bibfnamefont {M.~F.~M.}\ \bibnamefont {Lutz}},\ }\bibinfo {title} {{On Heavy light meson resonances and chiral symmetry}},\ \href {https://doi.org/10.1016/j.physletb.2003.10.118} {\bibfield  {journal} {\bibinfo  {journal} {Phys. Lett. B}\ }\textbf {\bibinfo {volume} {582}},\ \bibinfo {pages} {39} (\bibinfo {year} {2004})},\ \Eprint {https://arxiv.org/abs/hep-ph/0307133} {arXiv:hep-ph/0307133} \BibitemShut {NoStop}%
\bibitem [{\citenamefont {Gamermann}\ {\it et~al.}(2007)\citenamefont {Gamermann}, \citenamefont {Oset}, \citenamefont {Strottman},\ and\ \citenamefont {Vicente~Vacas}}]{Gamermann:2006nm}%
  \BibitemOpen
  \bibfield  {author} {\bibinfo {author} {\bibfnamefont {D.}~\bibnamefont {Gamermann}}, \bibinfo {author} {\bibfnamefont {E.}~\bibnamefont {Oset}}, \bibinfo {author} {\bibfnamefont {D.}~\bibnamefont {Strottman}},\ and\ \bibinfo {author} {\bibfnamefont {M.~J.}\ \bibnamefont {Vicente~Vacas}},\ }\bibinfo {title} {{Dynamically generated open and hidden charm meson systems}},\ \href {https://doi.org/10.1103/PhysRevD.76.074016} {\bibfield  {journal} {\bibinfo  {journal} {Phys. Rev. D}\ }\textbf {\bibinfo {volume} {76}},\ \bibinfo {pages} {074016} (\bibinfo {year} {2007})},\ \Eprint {https://arxiv.org/abs/hep-ph/0612179} {arXiv:hep-ph/0612179} \BibitemShut {NoStop}%
\bibitem [{\citenamefont {Guo}\ {\it et~al.}(2007)\citenamefont {Guo}, \citenamefont {Shen},\ and\ \citenamefont {Chiang}}]{Guo:2006rp}%
  \BibitemOpen
  \bibfield  {author} {\bibinfo {author} {\bibfnamefont {F.-K.}\ \bibnamefont {Guo}}, \bibinfo {author} {\bibfnamefont {P.-N.}\ \bibnamefont {Shen}},\ and\ \bibinfo {author} {\bibfnamefont {H.-C.}\ \bibnamefont {Chiang}},\ }\bibinfo {title} {{Dynamically generated $1^+$ heavy mesons}},\ \href {https://doi.org/10.1016/j.physletb.2007.01.050} {\bibfield  {journal} {\bibinfo  {journal} {Phys. Lett. B}\ }\textbf {\bibinfo {volume} {647}},\ \bibinfo {pages} {133} (\bibinfo {year} {2007})},\ \Eprint {https://arxiv.org/abs/hep-ph/0610008} {arXiv:hep-ph/0610008} \BibitemShut {NoStop}%
\bibitem [{\citenamefont {Guo}\ {\it et~al.}(2006)\citenamefont {Guo}, \citenamefont {Shen}, \citenamefont {Chiang}, \citenamefont {Ping},\ and\ \citenamefont {Zou}}]{Guo:2006fu}%
  \BibitemOpen
  \bibfield  {author} {\bibinfo {author} {\bibfnamefont {F.-K.}\ \bibnamefont {Guo}}, \bibinfo {author} {\bibfnamefont {P.-N.}\ \bibnamefont {Shen}}, \bibinfo {author} {\bibfnamefont {H.-C.}\ \bibnamefont {Chiang}}, \bibinfo {author} {\bibfnamefont {R.-G.}\ \bibnamefont {Ping}},\ and\ \bibinfo {author} {\bibfnamefont {B.-S.}\ \bibnamefont {Zou}},\ }\bibinfo {title} {{Dynamically generated $0^+$ heavy mesons in a heavy chiral unitary approach}},\ \href {https://doi.org/10.1016/j.physletb.2006.08.064} {\bibfield  {journal} {\bibinfo  {journal} {Phys. Lett. B}\ }\textbf {\bibinfo {volume} {641}},\ \bibinfo {pages} {278} (\bibinfo {year} {2006})},\ \Eprint {https://arxiv.org/abs/hep-ph/0603072} {arXiv:hep-ph/0603072} \BibitemShut {NoStop}%
\bibitem [{\citenamefont {Yang}\ {\it et~al.}(2022)\citenamefont {Yang}, \citenamefont {Wang}, \citenamefont {Wu}, \citenamefont {Oka},\ and\ \citenamefont {Zhu}}]{Yang:2021tvc}%
  \BibitemOpen
  \bibfield  {author} {\bibinfo {author} {\bibfnamefont {Z.}~\bibnamefont {Yang}}, \bibinfo {author} {\bibfnamefont {G.-J.}\ \bibnamefont {Wang}}, \bibinfo {author} {\bibfnamefont {J.-J.}\ \bibnamefont {Wu}}, \bibinfo {author} {\bibfnamefont {M.}~\bibnamefont {Oka}},\ and\ \bibinfo {author} {\bibfnamefont {S.-L.}\ \bibnamefont {Zhu}},\ }\bibinfo {title} {{Novel Coupled Channel Framework Connecting the Quark Model and Lattice QCD for the Near-threshold $D_s$ States}},\ \href {https://doi.org/10.1103/PhysRevLett.128.112001} {\bibfield  {journal} {\bibinfo  {journal} {Phys. Rev. Lett.}\ }\textbf {\bibinfo {volume} {128}},\ \bibinfo {pages} {112001} (\bibinfo {year} {2022})},\ \Eprint {https://arxiv.org/abs/2107.04860} {arXiv:2107.04860 [hep-ph]} \BibitemShut {NoStop}%
\bibitem [{\citenamefont {Liu}\ {\it et~al.}(2022)\citenamefont {Liu}, \citenamefont {Ling}, \citenamefont {Geng}, \citenamefont {Wang},\ and\ \citenamefont {Xie}}]{Liu:2022dmm}%
  \BibitemOpen
  \bibfield  {author} {\bibinfo {author} {\bibfnamefont {M.-Z.}\ \bibnamefont {Liu}}, \bibinfo {author} {\bibfnamefont {X.-Z.}\ \bibnamefont {Ling}}, \bibinfo {author} {\bibfnamefont {L.-S.}\ \bibnamefont {Geng}}, \bibinfo {author} {\bibfnamefont {E.}~\bibnamefont {Wang}},\ and\ \bibinfo {author} {\bibfnamefont {J.-J.}\ \bibnamefont {Xie}},\ }\bibinfo {title} {{Production of $D_{s0}^*(2317)$ and $D_{s1}(2460)$ in $B$ decays as $D^{(*)}K$ and $D_s$(*)\ensuremath{\eta} molecules}},\ \href {https://doi.org/10.1103/PhysRevD.106.114011} {\bibfield  {journal} {\bibinfo  {journal} {Phys. Rev. D}\ }\textbf {\bibinfo {volume} {106}},\ \bibinfo {pages} {114011} (\bibinfo {year} {2022})},\ \Eprint {https://arxiv.org/abs/2209.01103} {arXiv:2209.01103 [hep-ph]} \BibitemShut {NoStop}%
\bibitem [{\citenamefont {Godfrey}(2003)}]{Godfrey:2003kg}%
  \BibitemOpen
  \bibfield  {author} {\bibinfo {author} {\bibfnamefont {S.}~\bibnamefont {Godfrey}},\ }\bibinfo {title} {{Testing the nature of the $D_{sJ}^*(2317)^+$ and $D_{sJ}(2463)^+$ states using radiative transitions}},\ \href {https://doi.org/10.1016/j.physletb.2003.06.049} {\bibfield  {journal} {\bibinfo  {journal} {Phys. Lett. B}\ }\textbf {\bibinfo {volume} {568}},\ \bibinfo {pages} {254} (\bibinfo {year} {2003})},\ \Eprint {https://arxiv.org/abs/hep-ph/0305122} {arXiv:hep-ph/0305122} \BibitemShut {NoStop}%
\bibitem [{\citenamefont {Colangelo}\ and\ \citenamefont {De~Fazio}(2003)}]{Colangelo:2003vg}%
  \BibitemOpen
  \bibfield  {author} {\bibinfo {author} {\bibfnamefont {P.}~\bibnamefont {Colangelo}}\ and\ \bibinfo {author} {\bibfnamefont {F.}~\bibnamefont {De~Fazio}},\ }\bibinfo {title} {{Understanding $D_{sJ}(2317)$}},\ \href {https://doi.org/10.1016/j.physletb.2003.08.003} {\bibfield  {journal} {\bibinfo  {journal} {Phys. Lett. B}\ }\textbf {\bibinfo {volume} {570}},\ \bibinfo {pages} {180} (\bibinfo {year} {2003})},\ \Eprint {https://arxiv.org/abs/hep-ph/0305140} {arXiv:hep-ph/0305140} \BibitemShut {NoStop}%
\bibitem [{\citenamefont {Dmitrasinovic}(2005)}]{Dmitrasinovic:2005gc}%
  \BibitemOpen
  \bibfield  {author} {\bibinfo {author} {\bibfnamefont {V.}~\bibnamefont {Dmitrasinovic}},\ }\bibinfo {title} {{$D_{s0}^+(2317)$-$D_0(2308)$ mass difference as evidence for tetraquarks}},\ \href {https://doi.org/10.1103/PhysRevLett.94.162002} {\bibfield  {journal} {\bibinfo  {journal} {Phys. Rev. Lett.}\ }\textbf {\bibinfo {volume} {94}},\ \bibinfo {pages} {162002} (\bibinfo {year} {2005})}\BibitemShut {NoStop}%
\bibitem [{\citenamefont {Dmitrasinovic}(2012)}]{Dmitrasinovic:2012zz}%
  \BibitemOpen
  \bibfield  {author} {\bibinfo {author} {\bibfnamefont {V.}~\bibnamefont {Dmitrasinovic}},\ }\bibinfo {title} {{Chiral symmetry of heavy-light scalar mesons with $U_A(1)$ symmetry breaking}},\ \href {https://doi.org/10.1103/PhysRevD.86.016006} {\bibfield  {journal} {\bibinfo  {journal} {Phys. Rev. D}\ }\textbf {\bibinfo {volume} {86}},\ \bibinfo {pages} {016006} (\bibinfo {year} {2012})}\BibitemShut {NoStop}%
\bibitem [{\citenamefont {Colangelo}\ {\it et~al.}(2012)\citenamefont {Colangelo}, \citenamefont {De~Fazio}, \citenamefont {Giannuzzi},\ and\ \citenamefont {Nicotri}}]{Colangelo:2012xi}%
  \BibitemOpen
  \bibfield  {author} {\bibinfo {author} {\bibfnamefont {P.}~\bibnamefont {Colangelo}}, \bibinfo {author} {\bibfnamefont {F.}~\bibnamefont {De~Fazio}}, \bibinfo {author} {\bibfnamefont {F.}~\bibnamefont {Giannuzzi}},\ and\ \bibinfo {author} {\bibfnamefont {S.}~\bibnamefont {Nicotri}},\ }\bibinfo {title} {{New meson spectroscopy with open charm and beauty}},\ \href {https://doi.org/10.1103/PhysRevD.86.054024} {\bibfield  {journal} {\bibinfo  {journal} {Phys. Rev. D}\ }\textbf {\bibinfo {volume} {86}},\ \bibinfo {pages} {054024} (\bibinfo {year} {2012})},\ \Eprint {https://arxiv.org/abs/1207.6940} {arXiv:1207.6940 [hep-ph]} \BibitemShut {NoStop}%
\bibitem [{\citenamefont {Guo}(2023)}]{Guo:2023wkv}%
  \BibitemOpen
  \bibfield  {author} {\bibinfo {author} {\bibfnamefont {F.-K.}\ \bibnamefont {Guo}},\ }\bibinfo {title} {{Exotic hadrons from an effective field theory perspective}},\ \href {https://doi.org/10.22323/1.430.0232} {\bibfield  {journal} {\bibinfo  {journal} {PoS}\ }\textbf {\bibinfo {volume} {LATTICE2022}},\ \bibinfo {pages} {232} (\bibinfo {year} {2023})}\BibitemShut {NoStop}%
\bibitem [{\citenamefont {Liu}\ {\it et~al.}(2013)\citenamefont {Liu}, \citenamefont {Orginos}, \citenamefont {Guo}, \citenamefont {Hanhart},\ and\ \citenamefont {Mei{\ss}ner}}]{Liu:2012zya}%
  \BibitemOpen
  \bibfield  {author} {\bibinfo {author} {\bibfnamefont {L.}~\bibnamefont {Liu}}, \bibinfo {author} {\bibfnamefont {K.}~\bibnamefont {Orginos}}, \bibinfo {author} {\bibfnamefont {F.-K.}\ \bibnamefont {Guo}}, \bibinfo {author} {\bibfnamefont {C.}~\bibnamefont {Hanhart}},\ and\ \bibinfo {author} {\bibfnamefont {U.-G.}\ \bibnamefont {Mei{\ss}ner}},\ }\bibinfo {title} {{Interactions of charmed mesons with light pseudoscalar mesons from lattice QCD and implications on the nature of the $D_{s0}^*(2317)$}},\ \href {https://doi.org/10.1103/PhysRevD.87.014508} {\bibfield  {journal} {\bibinfo  {journal} {Phys. Rev. D}\ }\textbf {\bibinfo {volume} {87}},\ \bibinfo {pages} {014508} (\bibinfo {year} {2013})},\ \Eprint {https://arxiv.org/abs/1208.4535} {arXiv:1208.4535 [hep-lat]} \BibitemShut {NoStop}%
\bibitem [{\citenamefont {Mohler}\ {\it et~al.}(2013)\citenamefont {Mohler}, \citenamefont {Lang}, \citenamefont {Leskovec}, \citenamefont {Prelovsek},\ and\ \citenamefont {Woloshyn}}]{Mohler:2013rwa}%
  \BibitemOpen
  \bibfield  {author} {\bibinfo {author} {\bibfnamefont {D.}~\bibnamefont {Mohler}}, \bibinfo {author} {\bibfnamefont {C.~B.}\ \bibnamefont {Lang}}, \bibinfo {author} {\bibfnamefont {L.}~\bibnamefont {Leskovec}}, \bibinfo {author} {\bibfnamefont {S.}~\bibnamefont {Prelovsek}},\ and\ \bibinfo {author} {\bibfnamefont {R.~M.}\ \bibnamefont {Woloshyn}},\ }\bibinfo {title} {{$D_{s0}^*(2317)$ Meson and $D$-Meson-Kaon Scattering from Lattice QCD}},\ \href {https://doi.org/10.1103/PhysRevLett.111.222001} {\bibfield  {journal} {\bibinfo  {journal} {Phys. Rev. Lett.}\ }\textbf {\bibinfo {volume} {111}},\ \bibinfo {pages} {222001} (\bibinfo {year} {2013})},\ \Eprint {https://arxiv.org/abs/1308.3175} {arXiv:1308.3175 [hep-lat]} \BibitemShut {NoStop}%
\bibitem [{\citenamefont {Lang}\ {\it et~al.}(2014)\citenamefont {Lang}, \citenamefont {Leskovec}, \citenamefont {Mohler}, \citenamefont {Prelovsek},\ and\ \citenamefont {Woloshyn}}]{Lang:2014yfa}%
  \BibitemOpen
  \bibfield  {author} {\bibinfo {author} {\bibfnamefont {C.~B.}\ \bibnamefont {Lang}}, \bibinfo {author} {\bibfnamefont {L.}~\bibnamefont {Leskovec}}, \bibinfo {author} {\bibfnamefont {D.}~\bibnamefont {Mohler}}, \bibinfo {author} {\bibfnamefont {S.}~\bibnamefont {Prelovsek}},\ and\ \bibinfo {author} {\bibfnamefont {R.~M.}\ \bibnamefont {Woloshyn}},\ }\bibinfo {title} {{$D_s$ mesons with $DK$ and $D^*K$ scattering near threshold}},\ \href {https://doi.org/10.1103/PhysRevD.90.034510} {\bibfield  {journal} {\bibinfo  {journal} {Phys. Rev. D}\ }\textbf {\bibinfo {volume} {90}},\ \bibinfo {pages} {034510} (\bibinfo {year} {2014})},\ \Eprint {https://arxiv.org/abs/1403.8103} {arXiv:1403.8103 [hep-lat]} \BibitemShut {NoStop}%
\bibitem [{\citenamefont {Bali}\ {\it et~al.}(2017)\citenamefont {Bali}, \citenamefont {Collins}, \citenamefont {Cox},\ and\ \citenamefont {Sch\"afer}}]{Bali:2017pdv}%
  \BibitemOpen
  \bibfield  {author} {\bibinfo {author} {\bibfnamefont {G.~S.}\ \bibnamefont {Bali}}, \bibinfo {author} {\bibfnamefont {S.}~\bibnamefont {Collins}}, \bibinfo {author} {\bibfnamefont {A.}~\bibnamefont {Cox}},\ and\ \bibinfo {author} {\bibfnamefont {A.}~\bibnamefont {Sch\"afer}},\ }\bibinfo {title} {{Masses and decay constants of the $D_{s0}^*(2317)$ and $D_{s1}(2460)$ from $N_f=2$ lattice QCD close to the physical point}},\ \href {https://doi.org/10.1103/PhysRevD.96.074501} {\bibfield  {journal} {\bibinfo  {journal} {Phys. Rev. D}\ }\textbf {\bibinfo {volume} {96}},\ \bibinfo {pages} {074501} (\bibinfo {year} {2017})},\ \Eprint {https://arxiv.org/abs/1706.01247} {arXiv:1706.01247 [hep-lat]} \BibitemShut {NoStop}%
\bibitem [{\citenamefont {Cheung}\ {\it et~al.}(2021)\citenamefont {Cheung}, \citenamefont {Thomas}, \citenamefont {Wilson}, \citenamefont {Moir}, \citenamefont {Peardon},\ and\ \citenamefont {Ryan}}]{Cheung:2020mql}%
  \BibitemOpen
  \bibfield  {author} {\bibinfo {author} {\bibfnamefont {G.~K.~C.}\ \bibnamefont {Cheung}}, \bibinfo {author} {\bibfnamefont {C.~E.}\ \bibnamefont {Thomas}}, \bibinfo {author} {\bibfnamefont {D.~J.}\ \bibnamefont {Wilson}}, \bibinfo {author} {\bibfnamefont {G.}~\bibnamefont {Moir}}, \bibinfo {author} {\bibfnamefont {M.}~\bibnamefont {Peardon}},\ and\ \bibinfo {author} {\bibfnamefont {S.~M.}\ \bibnamefont {Ryan}} (\bibinfo {collaboration} {Hadron Spectrum}),\ }\bibinfo {title} {{$DK \;I = 0, D\bar{K} \; I = 0, 1$ scattering and the $ {D}_{s0}^{\ast } (2317)$ from lattice QCD}},\ \href {https://doi.org/10.1007/JHEP02(2021)100} {\bibfield  {journal} {\bibinfo  {journal} {JHEP}\ }\textbf {\bibinfo {volume} {02}},\ \bibinfo {pages} {100}},\ \Eprint {https://arxiv.org/abs/2008.06432} {arXiv:2008.06432 [hep-lat]} \BibitemShut {NoStop}%
\bibitem [{\citenamefont {Mart\'\i{}nez~Torres}\ {\it et~al.}(2015)\citenamefont {Mart\'\i{}nez~Torres}, \citenamefont {Oset}, \citenamefont {Prelovsek},\ and\ \citenamefont {Ramos}}]{MartinezTorres:2014kpc}%
  \BibitemOpen
  \bibfield  {author} {\bibinfo {author} {\bibfnamefont {A.}~\bibnamefont {Mart\'\i{}nez~Torres}}, \bibinfo {author} {\bibfnamefont {E.}~\bibnamefont {Oset}}, \bibinfo {author} {\bibfnamefont {S.}~\bibnamefont {Prelovsek}},\ and\ \bibinfo {author} {\bibfnamefont {A.}~\bibnamefont {Ramos}},\ }\bibinfo {title} {{Reanalysis of lattice QCD spectra leading to the $D_{s0}^*(2317)$ and $D_{s1}^*(2460)$}},\ \href {https://doi.org/10.1007/JHEP05(2015)153} {\bibfield  {journal} {\bibinfo  {journal} {JHEP}\ }\textbf {\bibinfo {volume} {05}},\ \bibinfo {pages} {153}},\ \Eprint {https://arxiv.org/abs/1412.1706} {arXiv:1412.1706 [hep-lat]} \BibitemShut {NoStop}%
\bibitem [{\citenamefont {Aaij}\ {\it et~al.}(2024)\citenamefont {Aaij} {\it et~al.}}]{LHCb:2023eeb}%
  \BibitemOpen
  \bibfield  {author} {\bibinfo {author} {\bibfnamefont {R.}~\bibnamefont {Aaij}} {\it et~al.} (\bibinfo {collaboration} {LHCb}),\ }\bibinfo {title} {{Observation of ${{\varLambda } ^0_{b}} \!\rightarrow {{\varLambda } ^+_{c}} {\hspace{1.79993pt}\overline{\hspace{-1.79993pt}D}} {}^{(*)0}{{K} ^-} $ and ${{\varLambda } ^0_{b}} \!\rightarrow {{\varLambda } ^+_{c}} {{D} ^{*-}_{s}} $ decays}},\ \href {https://doi.org/10.1140/epjc/s10052-024-12752-3} {\bibfield  {journal} {\bibinfo  {journal} {Eur. Phys. J. C}\ }\textbf {\bibinfo {volume} {84}},\ \bibinfo {pages} {575} (\bibinfo {year} {2024})},\ \Eprint {https://arxiv.org/abs/2311.14088} {arXiv:2311.14088 [hep-ex]} \BibitemShut {NoStop}%
\bibitem [{\citenamefont {Bando}\ {\it et~al.}(1988)\citenamefont {Bando}, \citenamefont {Kugo},\ and\ \citenamefont {Yamawaki}}]{Bando:1987br}%
  \BibitemOpen
  \bibfield  {author} {\bibinfo {author} {\bibfnamefont {M.}~\bibnamefont {Bando}}, \bibinfo {author} {\bibfnamefont {T.}~\bibnamefont {Kugo}},\ and\ \bibinfo {author} {\bibfnamefont {K.}~\bibnamefont {Yamawaki}},\ }\bibinfo {title} {{Nonlinear Realization and Hidden Local Symmetries}},\ \href {https://doi.org/10.1016/0370-1573(88)90019-1} {\bibfield  {journal} {\bibinfo  {journal} {Phys. Rept.}\ }\textbf {\bibinfo {volume} {164}},\ \bibinfo {pages} {217} (\bibinfo {year} {1988})}\BibitemShut {NoStop}%
\bibitem [{\citenamefont {Harada}\ and\ \citenamefont {Yamawaki}(2003)}]{Harada:2003jx}%
  \BibitemOpen
  \bibfield  {author} {\bibinfo {author} {\bibfnamefont {M.}~\bibnamefont {Harada}}\ and\ \bibinfo {author} {\bibfnamefont {K.}~\bibnamefont {Yamawaki}},\ }\bibinfo {title} {{Hidden local symmetry at loop: A New perspective of composite gauge boson and chiral phase transition}},\ \href {https://doi.org/10.1016/S0370-1573(03)00139-X} {\bibfield  {journal} {\bibinfo  {journal} {Phys. Rept.}\ }\textbf {\bibinfo {volume} {381}},\ \bibinfo {pages} {1} (\bibinfo {year} {2003})},\ \Eprint {https://arxiv.org/abs/hep-ph/0302103} {arXiv:hep-ph/0302103} \BibitemShut {NoStop}%
\bibitem [{\citenamefont {Mei{\ss}ner}(1988)}]{Meissner:1987ge}%
  \BibitemOpen
  \bibfield  {author} {\bibinfo {author} {\bibfnamefont {U.-G.}\ \bibnamefont {Mei{\ss}ner}},\ }\bibinfo {title} {{Low-Energy Hadron Physics from Effective Chiral Lagrangians with Vector Mesons}},\ \href {https://doi.org/10.1016/0370-1573(88)90090-7} {\bibfield  {journal} {\bibinfo  {journal} {Phys. Rept.}\ }\textbf {\bibinfo {volume} {161}},\ \bibinfo {pages} {213} (\bibinfo {year} {1988})}\BibitemShut {NoStop}%
\bibitem [{\citenamefont {Nagahiro}\ {\it et~al.}(2009)\citenamefont {Nagahiro}, \citenamefont {Roca}, \citenamefont {Hosaka},\ and\ \citenamefont {Oset}}]{Nagahiro:2008cv}%
  \BibitemOpen
  \bibfield  {author} {\bibinfo {author} {\bibfnamefont {H.}~\bibnamefont {Nagahiro}}, \bibinfo {author} {\bibfnamefont {L.}~\bibnamefont {Roca}}, \bibinfo {author} {\bibfnamefont {A.}~\bibnamefont {Hosaka}},\ and\ \bibinfo {author} {\bibfnamefont {E.}~\bibnamefont {Oset}},\ }\bibinfo {title} {{Hidden gauge formalism for the radiative decays of axial-vector mesons}},\ \href {https://doi.org/10.1103/PhysRevD.79.014015} {\bibfield  {journal} {\bibinfo  {journal} {Phys. Rev. D}\ }\textbf {\bibinfo {volume} {79}},\ \bibinfo {pages} {014015} (\bibinfo {year} {2009})},\ \Eprint {https://arxiv.org/abs/0809.0943} {arXiv:0809.0943 [hep-ph]} \BibitemShut {NoStop}%
\bibitem [{\citenamefont {Press}\ {\it et~al.}(1992)\citenamefont {Press}, \citenamefont {Teukolsky}, \citenamefont {Vetterling},\ and\ \citenamefont {Flannery}}]{Press1992}%
  \BibitemOpen
  \bibfield  {author} {\bibinfo {author} {\bibfnamefont {W.}~\bibnamefont {Press}}, \bibinfo {author} {\bibfnamefont {S.}~\bibnamefont {Teukolsky}}, \bibinfo {author} {\bibfnamefont {W.}~\bibnamefont {Vetterling}},\ and\ \bibinfo {author} {\bibfnamefont {B.}~\bibnamefont {Flannery}},\ }\href@noop {} {\bibinfo {title} {Numerical recipes in {FORTRAN}: The art of scientific computing}} (\bibinfo {year} {1992})\BibitemShut {NoStop}%
\bibitem [{\citenamefont {Efron}\ and\ \citenamefont {Tibshirani}(1986)}]{Efron:1986hys}%
  \BibitemOpen
  \bibfield  {author} {\bibinfo {author} {\bibfnamefont {B.}~\bibnamefont {Efron}}\ and\ \bibinfo {author} {\bibfnamefont {R.}~\bibnamefont {Tibshirani}},\ }\bibinfo {title} {{An introduction to the bootstrap}},\ \href@noop {} {\bibfield  {journal} {\bibinfo  {journal} {Statist. Sci.}\ }\textbf {\bibinfo {volume} {57}},\ \bibinfo {pages} {54} (\bibinfo {year} {1986})}\BibitemShut {NoStop}%
\bibitem [{\citenamefont {Albaladejo}\ {\it et~al.}(2016)\citenamefont {Albaladejo}, \citenamefont {Jido}, \citenamefont {Nieves},\ and\ \citenamefont {Oset}}]{Albaladejo:2016hae}%
  \BibitemOpen
  \bibfield  {author} {\bibinfo {author} {\bibfnamefont {M.}~\bibnamefont {Albaladejo}}, \bibinfo {author} {\bibfnamefont {D.}~\bibnamefont {Jido}}, \bibinfo {author} {\bibfnamefont {J.}~\bibnamefont {Nieves}},\ and\ \bibinfo {author} {\bibfnamefont {E.}~\bibnamefont {Oset}},\ }\bibinfo {title} {{$D^*_{s0}(2317)$ and ${DK}$ scattering in $B$ decays from BaBar and LHCb data}},\ \href {https://doi.org/10.1140/epjc/s10052-016-4144-3} {\bibfield  {journal} {\bibinfo  {journal} {Eur. Phys. J. C}\ }\textbf {\bibinfo {volume} {76}},\ \bibinfo {pages} {300} (\bibinfo {year} {2016})},\ \Eprint {https://arxiv.org/abs/1604.01193} {arXiv:1604.01193 [hep-ph]} \BibitemShut {NoStop}%
\bibitem [{\citenamefont {Ikeno}\ {\it et~al.}(2023)\citenamefont {Ikeno}, \citenamefont {Toledo},\ and\ \citenamefont {Oset}}]{Ikeno:2023ojl}%
  \BibitemOpen
  \bibfield  {author} {\bibinfo {author} {\bibfnamefont {N.}~\bibnamefont {Ikeno}}, \bibinfo {author} {\bibfnamefont {G.}~\bibnamefont {Toledo}},\ and\ \bibinfo {author} {\bibfnamefont {E.}~\bibnamefont {Oset}},\ }\bibinfo {title} {{Model independent analysis of femtoscopic correlation functions: An application to the $D_{s0}^*(2317)$}},\ \href {https://doi.org/10.1016/j.physletb.2023.138281} {\bibfield  {journal} {\bibinfo  {journal} {Phys. Lett. B}\ }\textbf {\bibinfo {volume} {847}},\ \bibinfo {pages} {138281} (\bibinfo {year} {2023})},\ \Eprint {https://arxiv.org/abs/2305.16431} {arXiv:2305.16431 [hep-ph]} \BibitemShut {NoStop}%
\bibitem [{\citenamefont {Gamermann}\ {\it et~al.}(2010)\citenamefont {Gamermann}, \citenamefont {Nieves}, \citenamefont {Oset},\ and\ \citenamefont {Ruiz~Arriola}}]{Gamermann:2009uq}%
  \BibitemOpen
  \bibfield  {author} {\bibinfo {author} {\bibfnamefont {D.}~\bibnamefont {Gamermann}}, \bibinfo {author} {\bibfnamefont {J.}~\bibnamefont {Nieves}}, \bibinfo {author} {\bibfnamefont {E.}~\bibnamefont {Oset}},\ and\ \bibinfo {author} {\bibfnamefont {E.}~\bibnamefont {Ruiz~Arriola}},\ }\bibinfo {title} {{Couplings in coupled channels versus wave functions: application to the $X(3872)$ resonance}},\ \href {https://doi.org/10.1103/PhysRevD.81.014029} {\bibfield  {journal} {\bibinfo  {journal} {Phys. Rev. D}\ }\textbf {\bibinfo {volume} {81}},\ \bibinfo {pages} {014029} (\bibinfo {year} {2010})},\ \Eprint {https://arxiv.org/abs/0911.4407} {arXiv:0911.4407 [hep-ph]} \BibitemShut {NoStop}%
\bibitem [{\citenamefont {Song}\ {\it et~al.}(2022)\citenamefont {Song}, \citenamefont {Dai},\ and\ \citenamefont {Oset}}]{Song:2022yvz}%
  \BibitemOpen
  \bibfield  {author} {\bibinfo {author} {\bibfnamefont {J.}~\bibnamefont {Song}}, \bibinfo {author} {\bibfnamefont {L.~R.}\ \bibnamefont {Dai}},\ and\ \bibinfo {author} {\bibfnamefont {E.}~\bibnamefont {Oset}},\ }\bibinfo {title} {{How much is the compositeness of a bound state constrained by $a$ and $r_0$? The role of the interaction range}},\ \href {https://doi.org/10.1140/epja/s10050-022-00753-3} {\bibfield  {journal} {\bibinfo  {journal} {Eur. Phys. J. A}\ }\textbf {\bibinfo {volume} {58}},\ \bibinfo {pages} {133} (\bibinfo {year} {2022})},\ \Eprint {https://arxiv.org/abs/2201.04414} {arXiv:2201.04414 [hep-ph]} \BibitemShut {NoStop}%
\bibitem [{\citenamefont {Hyodo}(2013)}]{Hyodo:2013nka}%
  \BibitemOpen
  \bibfield  {author} {\bibinfo {author} {\bibfnamefont {T.}~\bibnamefont {Hyodo}},\ }\bibinfo {title} {{Structure and compositeness of hadron resonances}},\ \href {https://doi.org/10.1142/S0217751X13300457} {\bibfield  {journal} {\bibinfo  {journal} {Int. J. Mod. Phys. A}\ }\textbf {\bibinfo {volume} {28}},\ \bibinfo {pages} {1330045} (\bibinfo {year} {2013})},\ \Eprint {https://arxiv.org/abs/1310.1176} {arXiv:1310.1176 [hep-ph]} \BibitemShut {NoStop}%
\bibitem [{\citenamefont {Hyodo}\ {\it et~al.}(2012)\citenamefont {Hyodo}, \citenamefont {Jido},\ and\ \citenamefont {Hosaka}}]{Hyodo:2011qc}%
  \BibitemOpen
  \bibfield  {author} {\bibinfo {author} {\bibfnamefont {T.}~\bibnamefont {Hyodo}}, \bibinfo {author} {\bibfnamefont {D.}~\bibnamefont {Jido}},\ and\ \bibinfo {author} {\bibfnamefont {A.}~\bibnamefont {Hosaka}},\ }\bibinfo {title} {{Compositeness of dynamically generated states in a chiral unitary approach}},\ \href {https://doi.org/10.1103/PhysRevC.85.015201} {\bibfield  {journal} {\bibinfo  {journal} {Phys. Rev. C}\ }\textbf {\bibinfo {volume} {85}},\ \bibinfo {pages} {015201} (\bibinfo {year} {2012})},\ \Eprint {https://arxiv.org/abs/1108.5524} {arXiv:1108.5524 [nucl-th]} \BibitemShut {NoStop}%
\bibitem [{\citenamefont {Liang}\ {\it et~al.}(2014)\citenamefont {Liang}, \citenamefont {Xiao},\ and\ \citenamefont {Oset}}]{Liang:2014eba}%
  \BibitemOpen
  \bibfield  {author} {\bibinfo {author} {\bibfnamefont {W.~H.}\ \bibnamefont {Liang}}, \bibinfo {author} {\bibfnamefont {C.~W.}\ \bibnamefont {Xiao}},\ and\ \bibinfo {author} {\bibfnamefont {E.}~\bibnamefont {Oset}},\ }\bibinfo {title} {{Baryon states with open beauty in the extended local hidden gauge approach}},\ \href {https://doi.org/10.1103/PhysRevD.89.054023} {\bibfield  {journal} {\bibinfo  {journal} {Phys. Rev. D}\ }\textbf {\bibinfo {volume} {89}},\ \bibinfo {pages} {054023} (\bibinfo {year} {2014})},\ \Eprint {https://arxiv.org/abs/1401.1441} {arXiv:1401.1441 [hep-ph]} \BibitemShut {NoStop}%
\bibitem [{\citenamefont {Dias}\ {\it et~al.}(2020)\citenamefont {Dias}, \citenamefont {Yu}, \citenamefont {Liang}, \citenamefont {Sun}, \citenamefont {Xie},\ and\ \citenamefont {Oset}}]{Dias:2019klk}%
  \BibitemOpen
  \bibfield  {author} {\bibinfo {author} {\bibfnamefont {J.~M.}\ \bibnamefont {Dias}}, \bibinfo {author} {\bibfnamefont {Q.-X.}\ \bibnamefont {Yu}}, \bibinfo {author} {\bibfnamefont {W.-H.}\ \bibnamefont {Liang}}, \bibinfo {author} {\bibfnamefont {Z.-F.}\ \bibnamefont {Sun}}, \bibinfo {author} {\bibfnamefont {J.-J.}\ \bibnamefont {Xie}},\ and\ \bibinfo {author} {\bibfnamefont {E.}~\bibnamefont {Oset}},\ }\bibinfo {title} {{$\Xi_{bb}$ and $\Omega_{bbb}$ molecular states}},\ \href {https://doi.org/10.1088/1674-1137/44/6/064101} {\bibfield  {journal} {\bibinfo  {journal} {Chin. Phys. C}\ }\textbf {\bibinfo {volume} {44}},\ \bibinfo {pages} {064101} (\bibinfo {year} {2020})},\ \Eprint {https://arxiv.org/abs/1912.04517} {arXiv:1912.04517 [hep-ph]} \BibitemShut {NoStop}%
\bibitem [{\citenamefont {Aceti}\ {\it et~al.}(2014)\citenamefont {Aceti}, \citenamefont {Dai}, \citenamefont {Geng}, \citenamefont {Oset},\ and\ \citenamefont {Zhang}}]{Aceti:2014ala}%
  \BibitemOpen
  \bibfield  {author} {\bibinfo {author} {\bibfnamefont {F.}~\bibnamefont {Aceti}}, \bibinfo {author} {\bibfnamefont {L.~R.}\ \bibnamefont {Dai}}, \bibinfo {author} {\bibfnamefont {L.~S.}\ \bibnamefont {Geng}}, \bibinfo {author} {\bibfnamefont {E.}~\bibnamefont {Oset}},\ and\ \bibinfo {author} {\bibfnamefont {Y.}~\bibnamefont {Zhang}},\ }\bibinfo {title} {{Meson-baryon components in the states of the baryon decuplet}},\ \href {https://doi.org/10.1140/epja/i2014-14057-2} {\bibfield  {journal} {\bibinfo  {journal} {Eur. Phys. J. A}\ }\textbf {\bibinfo {volume} {50}},\ \bibinfo {pages} {57} (\bibinfo {year} {2014})},\ \Eprint {https://arxiv.org/abs/1301.2554} {arXiv:1301.2554 [hep-ph]} \BibitemShut {NoStop}%
\end{thebibliography}%
\end{document}